\documentclass[lettersize,journal]{IEEEtran}
\usepackage{amsmath,amsfonts,amssymb}
\usepackage{amsthm,bbold}
\usepackage{empheq}
\usepackage{algorithm,algpseudocode}
\usepackage{array}
\usepackage{subcaption}
\usepackage{textcomp}
\usepackage{tcolorbox}
\usepackage{stfloats}
\usepackage{url}
\usepackage{verbatim}
\usepackage{graphicx}
\usepackage{cite}
\usepackage{tabularx}%
\usepackage{graphicx,booktabs, colortbl}
\usepackage{multirow}
\usepackage{soul}
\usepackage{xcolor,soul}

\newtheorem{theorem}{Theorem}

\newtheorem{definition}{Definition}

\usepackage{bm}

\newcommand*{\etaltwo}{\textit{et al}.\@ }

\begin{document}

\title{Cascade-driven opinion dynamics\\on social networks}

\author{Elisabetta Biondi, Chiara Boldrini, Andrea Passarella, Marco Conti \vspace{-1cm}
\thanks{All authors are with the Institute of Informatics and Telematics (IIT) of the National Research Council (CNR), Italy. email: first.last@iit.cnr.it}
\thanks{The work of M. Conti and A. Passarella is partly supported by PNRR - M4C2 - Investimento 1.3, Partenariato Esteso PE00000013 - ``FAIR - Future Artificial Intelligence Research'' - Spoke 1 "Human-centered AI", funded by the European Commission under the NextGeneration EU programme. 
The work of C. Boldrini and E. Biondi is partly supported by project SERICS (PE00000014) under the MUR National Recovery and Resilience Plan funded by the European Union - NextGenerationEU.
This work is also supported by the European Union under the scheme HORIZON-INFRA-2021-DEV-02-01 – Preparatory phase of new ESFRI research infrastructure projects, Grant Agreement n.101079043, “SoBigData RI PPP: SoBigData RI Preparatory Phase Project”. 
This work was partially supported by SoBigData.it. SoBigData.it receives funding from European Union – NextGenerationEU – National Recovery and Resilience Plan (Piano Nazionale di Ripresa e Resilienza, PNRR) – Project: “SoBigData.it – Strengthening the Italian RI for Social Mining and Big Data Analytics” – Prot. IR0000013 – Avviso n. 3264 del 28/12/2021. }
}

\markboth{Journal of \LaTeX\ Class Files,~Vol.~14, No.~8, August~2021}%
{Shell \MakeLowercase{\textit{et al.}}: A Sample Article Using IEEEtran.cls for IEEE Journals}


\maketitle


\begin{abstract}
Online social networks (OSNs) have transformed the way individuals fulfill their social needs and consume information. As OSNs become increasingly prominent sources for news dissemination, individuals often encounter content that influences their opinions through both direct interactions and broader network dynamics. In this paper, we propose the Friedkin-Johnsen on Cascade (FJC) model, which, to the best of our knowledge, is the first attempt to integrate information cascades and opinion dynamics, specifically using the very popular Friedkin-Johnsen model. Our model, validated over real social cascades, highlights how the convergence of socialization and sharing news on these platforms can disrupt opinion evolution dynamics typically observed in offline settings. Our findings demonstrate that these cascades can amplify the influence of central opinion leaders, making them more resistant to divergent viewpoints, even when challenged by a critical mass of dissenting opinions. This research underscores the importance of understanding the interplay between social dynamics and information flow in shaping public discourse in the digital age.
\end{abstract}

\begin{IEEEkeywords}
Information cascades, opinion dynamics, polarization, Friedkin-Johnsen model, viral content
\end{IEEEkeywords}

\section{Introduction}
\IEEEPARstart{O}{nline} social networks (OSNs) have fundamentally transformed how individuals communicate, share information, and form opinions in the digital age. Serving both as platforms for social interaction and as primary sources of real-time news, OSNs facilitate unprecedented levels of connectivity and information dissemination~\cite{matsa2018news}.
Moreover, media organizations have implemented strategies to encourage the sharing of their content on OSNs, facilitating the seamless integration of news into users' social interactions. Therefore, during our social activities online, we can also be exposed to the news even incidentally 
raising concerns about filter bubbles, echo chambers, and the rapid spread of misinformation~\cite{moller2020explaining}. These phenomena highlight the need to understand how information diffusion and social influence jointly shape opinion formation in online environments.

The dynamics of opinion formation represent a multifaceted research area situated at the intersection of various disciplines, including sociology, economics, psychology, control theory, computer science, computational social science, and physics. Numerous models have been proposed to describe the mechanisms underlying opinion dynamics, aiming to capture observed theoretical and empirical phenomena. 
Among the prominent models in this domain is the Friedkin-Johnsen (FJ) model~\cite{friedkin1990social} belonging to the category of \emph{averaging models}~\cite{degroot1974reaching,french1956formal,friedkin1990social}, according to which people iteratively adjust their opinions towards the average opinion of their neighbours, with the strength of social ties weighting the influence of each neighbor's opinion.  Additionally, the FJ model incorporates a parameter representing individuals' open-mindedness, referred to as \emph{susceptibility}, which modulates the extent to which individuals revise their opinions. The FJ model has gained popularity due to its mathematical tractability and its empirical validation on real social networks. It has demonstrated the ability to replicate various observed mechanisms in social networks, including the phenomenon of wisdom of the crowd~\cite{Das2013} and the emergence of polarization~\cite{biondi2023dynamics} and recently, it has also been proposed as the building block to model multilateral international negotiation processes~\cite{bernardo2021achieving}. 

The Friedkin-Johnsen model has been validated in offline settings, where opinion evolution reflects direct interpersonal interactions and deliberations across various topics~\cite{Friedkin2011,Friedkin2017}. However, online social environments introduce distinct mechanisms of opinion exchange: individuals are exposed to their contacts’ expressed opinions indirectly, primarily through public social media activities such as posts, comments, and shares~\cite{garimella2018quantifying,suarez2025cross}. This mode of exposure frequently involves repeated encounters with viewpoints from a broader and more heterogeneous audience, which can heighten cognitive dissonance by presenting conflicting information~\cite{festinger1957theory,tsfati2003media}. In online environments, opinion exposure is both asynchronous and large-scale, allowing conflicting signals to endure over time and reach diverse audiences. At the same time, empirical studies show that controversial or polarizing opinions propagate with exceptional speed and breadth compared to offline interactions, amplifying their influence on collective attitudes~\cite{vosoughi2018spread,delvicario2016misinformation}. These opinion expressions propagate through various content formats, including textual posts, news resharing, and multimedia, whose spread dynamics are effectively modeled by the theory of \emph{information cascades}~\cite{easley2010networks}.


Opinion dynamics and information cascades operate at distinct granularity and temporal scales. Information cascades focus on how individual pieces of information (e.g., a post, an image) are shared and propagate within the social network, while opinion dynamics concern the evolution of opinions through interactions among individuals. Both processes exhibit emergent behaviors, transitioning from local interactions to global effects such as content virality and polarization. In opinion dynamics, the specific social interactions leading to opinion change remain abstract and unspecified. Conversely, information cascades emphasize the sharing mechanism of content without addressing the effects of such sharing. 
Importantly, reposting online content received from friends (e.g., retweets) is one of the most common forms of interaction in digital platforms. Empirical evidence shows that in a complete 24-hour snapshot of Twitter activity, only 20.8\% of tweets were original, while 50.7\% were retweets, underscoring that resharing dominates user behavior~\cite{pfeffer2023just}. This widespread resharing suggests a natural intertwining between opinion dynamics and information cascades: as users repeatedly encounter and propagate content, microscopic acts of resharing can influence macroscopic opinion evolution. Despite this evident overlap, the two processes have largely been treated separately in the literature. 
\textbf{In this work, we set out to fill this gap and establish an integrated model whereby the microscopic act of content sharing bears an effect in the macroscopic phenomenon of opinion evolution in online social networks. 
}


To address this issue, we propose the Friedkin–Johnsen on Cascades (FJC) model, a novel framework that explicitly couples opinion dynamics with information diffusion. FJC departs from the classical FJ model by replacing abstract, synchronous interpersonal interactions with empirically grounded cascade-based exposure: opinions are updated only when users receive reposted content through an information cascade. In this setting, the social influence exerted by a neighbour is activated by the resharing process itself and is therefore constrained by the temporal order, reach, and structure of cascades. 
Concretely, cascade sources initiate posts that propagate through the network according to resharing probabilities, and each resharing event conveys the sharer’s expressed opinion on the topic. Opinion updates thus occur asynchronously and selectively, driven by actual exposure rather than by static network adjacency. This mechanism links the microscopic dynamics of content diffusion to the macroscopic evolution of opinions, yielding an opinion-formation process that is inherently shaped by cascade depth, timing, and network position.


%
First, we validated our FJC model on a real dataset of information cascades from Twitter, demonstrating its strong predictive power for online opinion dynamics. Building upon our previous work~\cite{biondi2023dynamics}, we then analyzed the impact of cascade mechanisms on polarization, evaluating polarization levels using opinion vectors that maximize polarization in the classical FJ model. To this end, we tested the model on both synthetic and real social networks.  Using FJC, we systematically compare opinion evolution under cascade-driven exposure with the classical FJ dynamics, both on real cascade data and on controlled synthetic networks.
Our main findings can be summarized as follows:
\begin{itemize}
    \item Highly central nodes emerge as opinion leaders with disproportionate influence, capable of either intensifying or mitigating polarization. Notably, regardless of the extent of information spread, the sequential nature of cascades causes central nodes to encounter content early, limiting their exposure to diverse opinions and making them less susceptible to external influence.
    \item The FJC dynamics weaken the influence of peripheral nodes—even when they are numerous, stubborn, and aligned—since they often fail to reach and sway the central nodes.
    \item Information cascades can substantially shape polarization outcomes, either amplifying or reducing them, depending on how opinions spread through the network.
\end{itemize}
These findings underscore how sharing mechanisms on online platforms introduce structural and temporal asymmetries that reshape opinion evolution and amplify the role of network topology compared to the classical FJ model. Furthermore, they highlight how online dynamics can exacerbate polarization beyond what is typically observed in offline social contexts. 
Neglecting the role of diffusion dynamics can therefore lead to serious misjudgments in assessing both opinion formation and polarization. These insights highlight the critical role of online information diffusion in shaping collective attitudes and social interactions and offer guidance for communication strategies in business, public discourse, and policy-making.

\vspace{-5pt}
\section{Related Works}
\label{sec:relwork}

The modeling of opinion dynamics has attracted sustained interest for decades, an interest that has further intensified with the rise of online social networks. The literature is extensive and encompasses a broad spectrum of models of increasing complexity~\cite{peralta2025opinion}. 
Here, we concentrate on the class of \emph{averaging models}, which posit that users adjust their opinions based on those of the people around them to mitigate social discomfort.
Among these models, the FJ model~\cite{friedkin1990social} is one of the most studied~\cite{Gionis2013,Bindel2015,Matakos2018,Musco2018,Chen2018,proskurnikov2017tutorial,proskurnikov2018dynamics,Friedkin2017,Friedkin2015,biondi2023dynamics}. Its inner workings are explained in Section~\ref{sec:fj}.

Recently, research on the FJ model has taken two main directions. The first involves optimization problems based on FJ~\cite{Tu2023,Sun2023,abebe2021opinion,Musco2018,Gionis2013,Bindel2015}, whose main goal is to make interventions into the social network to achieve specific desired outcomes in the final opinions. For example, some investigate how to modify the initial opinion of a set of nodes to maximise the opinion discord~\cite{Tu2023} or minimize the overall opinion~\cite{Sun2023}. The second direction explores the possible extensions of the FJ model to capture diverse observed features in the online social networks~\cite{Bhalla2023,bilo2022general,chitra2019understanding,Razaq2023,disaro2023extension}. Several extensions of the FJ model incorporate repulsion~\cite{bilo2022general,Razaq2023}, homophily~\cite{disaro2023extension}, filter bubbles~\cite{chitra2019understanding}, and temporal networks~\cite{Bhalla2023}. 
Differently from these works, we investigate the effect of the classical FJ model on a static network when the social debate is shaped by the temporal flow of news posts on online social platforms. \\
Beyond FJ-specific models, similar distinctions appear in the broader opinion dynamics literature. On the one hand, continuous-time and kinetic frameworks~\cite{albi2015optimal, zanella2023kinetic} aim to capture network evolution and large-scale interaction effects. In continuous-time models, the underlying social network may evolve over time, while kinetic approaches model opinion interactions in a mean-field setting, often drawing analogies with epidemic spreading and allowing agents’ susceptibility to influence to depend endogenously on their opinions. On the other hand, recent research has emphasized prescriptive interventions aimed at mitigating polarization in social networks. For instance, Liu and Wen~\cite{liu2024agent} propose an agent-based framework that identifies network motifs to design targeted interventions by modifying the interaction structure. Within this landscape, the FJC model provides a foundational, cascade-aware description of how opinions evolve over time in online environments, which can naturally inform future intervention and application-oriented extensions.

To model the posting flow, we draw inspiration from the independent-cascade model in~\cite{kempe2003maximizing} (see Section~\ref{sec:information_cascade} for a description of how it works), which is able to reproduce realistic diffusion dynamics~\cite{guille2013information}. This seminal work has inspired many following studies and extensions~\cite{barbieri2013topic,aslay2016revenue,chen2011influence,gayraud2015diffusion}, which extend the classical model to include collaborative and/or antagonistic relationships among users, as well as dynamic networks. While information diffusion studies typically focus on content popularity, we are interested in how the diffusion process itself affects opinion formation. 
Hence, we used the original independent-cascade model and analyzed its impact on the opinion formation process. Similarly to opinion dynamics, the literature on information diffusion is extensive, so we refer the interested reader to dedicated surveys for more details~\cite{zhou2021survey}.

The works most closely related to ours are~\cite{tu2022viral,brooks2025opinion}, which also investigate the interplay between information diffusion and opinion formation. While~\cite{tu2022viral} adopts the Friedkin–Johnsen model and~\cite{brooks2025opinion} employs a bounded confidence framework, both approaches differ fundamentally from ours. 
In~\cite{brooks2025opinion}, opinion adoption is driven solely by similarity to the root node’s opinion and is primarily used to model content virality, not the cognitive process of opinion change. In contrast, our FJC model assumes that every act of content sharing conveys the sharer’s current opinion, which in turn directly influences the recipient’s belief. This enables us to explicitly capture how the temporal structure of cascades shapes long-term opinion dynamics and polarization—an aspect not addressed in prior work.
Tu and Neumann~\cite{tu2022viral}, instead, study the interaction between viral content and opinion dynamics. In their model, opinion evolution and information diffusion are treated as largely separate mechanisms. Opinion updates are governed by a relaxed version of the FJ model, while content diffusion affects opinions only through a one-time adjustment of individuals’ innate opinions, based on the final popularity of the content. After this initial bias shift, opinions evolve independently of further cascade activity. 
In contrast, in our framework, content diffusion and opinion dynamics are tightly coupled: sharing content is interpreted as an explicit expression of an individual’s opinion, and each exposure to reshared content can directly trigger an opinion update. As a result, information cascades continuously shape the opinion formation process over time, rather than acting only through an initial perturbation.

Complementary to our approach, Scarpa~\etaltwo~\cite{scarpa2025polarization} develop a Markovian agent-based framework for transient opinion dynamics in spatially structured offline settings. While their work emphasizes short-term and location-dependent effects, our model focuses on large-scale online cascades and emergent polarization phenomena.

\vspace{-5pt}
\section{Preliminaries}
\label{sec:preliminaries}

In this section, we will summarize the theoretical notions and notation we will use throughout the rest of the paper.

\vspace{-5pt}

\subsection{The Social Network} 
\label{sec:social_network}
We can describe an online social network platform as a graph $\mathcal{G}=(\mathcal{V},\mathcal{E})$, where the $v$ nodes $\mathcal{V}$ represent users and the edges $\mathcal{E}$ represent social connections between them. In the most general form, the graph $G$ is directed and the edges $(i,j)$ represent a \emph{follow} relationship. Specifically,  $(i,j)\in\mathcal{E}$ means that $i$ \emph{follows} $j$, thus $j$ influences $i$ as its activity appears in $i$'s news feed. The set of nodes that has an influence on $i$ is thus composed of $i$'s out-neighbours, which we will denote with $\mathcal{N}^+(i)$. Instead, the set of nodes that is influenced by node $j$ is the in-neighbourhood of $j$, which we indicate with $\mathcal{N}^-(j)$\footnote{Please note that, generally, the \emph{neighbourhood} of a node corresponds to our out-neighbourhood.}. It is worth noting that in the special case of an undirected node, $\mathcal{N}^+(i)$ and $\mathcal{N}^-(i)$ are the same.
This social connection can arise, for example, when $i$ follows $j$ on platforms like X or Facebook or when they are friends on Facebook. 
The strength of this relationship is mathematically represented by the quantity $\hat{w}_{ij}$, which is the $(i,j)$-th entry of the $n\times n$ matrix $\hat{W}$ that we call \emph{social matrix}. The weight that node $i$ gives to its out-neighbours represents the degree of influence of their opinions. It is provided by the $(i,j)$-th entry of the $n\times n$ matrix $W$, named the \emph{influence matrix} $W$, which is row-stochastic. On the $i$-th row, for example, we can see how the influence on $i$'s opinion is distributed among all nodes in $i$'s network. 
In situations where additional data is lacking, the matrix $W$ can be derived from $\hat{W}$ through row-normalization, expressed as $w_{ij} = \frac{\hat{w}{ij}}{\sum_j \hat{w}{ij}}$. This process capitalizes on the insight that stronger relationships should carry greater weight in opinion formation. It is worth noting that, with this mechanism, even in cases where the social graph exhibits symmetry, the influence matrix results asymmetric. 

\vspace{-5pt}
\subsection{Information Cascade}
\label{sec:information_cascade} 

Our model conceptualizes the act of posting social content following the independent cascade model~\cite{kempe2003maximizing}. In the independent cascade model, each node can be \emph{active} or \emph{inactive}. An initial set of active nodes $v$ has a single chance to activate their neighbours $w$, and each activation is successful with probability $p_{vw}$. The nodes $w$ that are successfully activated become active at the next step and try to activate their neighbours as well.  The process continues until no other nodes can be activated. This framework is well-suited to modeling the propagation of shared elements in online social networks, such as tweets or posts, where each user typically has one opportunity to reshare content upon seeing it, and further reshares depend on their direct followers. 
This cascade model forms the basis of the proposed FJC framework and is used extensively in Sections~\ref{sec:model_setting} and~\ref{sec:fjc}.

\vspace{-5pt}
\subsection{Friedkin-Johnsen Opinion Dynamics Model}
\label{sec:fj}

The FJ opinion dynamic model is a mathematical model that describes how people's opinions change over time due to their mutual influences. 
The FJ model is mathematically expressed as a simple update equation that is applied to each person's opinion at discrete times. As discussed in our previous work~\cite{biondi2023dynamics}, in the literature, the term ``FJ model'' refers to many variants of the original model. The update expression for the most general variant of the FJ model is given by the following equation: 
\begin{equation}
    x_i(t+1)=\lambda_i\sum_{j\in\mathcal{V}} w_{ij}x_j(t)+(1-\lambda_i)u_i\qquad \forall i,\label{eq:fj}
    \vspace{-5pt}
\end{equation}
where $ w_{ij}$ is the $(i,j)$-th entry of matrix $W$. The first term of the sum represents the movement of $i$'s opinion towards the opinion of its neighbours, weighted by their influence degree. The second term represents instead the anchoring to the initial prejudice. Under general assumptions, the model converges, as summarized in the following theorem (proved in~\cite{proskurnikov2017tutorial}).
\begin{theorem}\label{theo:fj_convergence}[FJ convergence] If the graph is strongly connected and $\Lambda \ne I$ (i.e., at least one agent partially anchored to its initial opinion exists)\footnote{Please note that when $\Lambda=I$, the FJ model reduces to the French-DeGroot model~\cite{degroot1974reaching}, which is a simpler model that has been separately studied in the literature.} then the FJ model is convergent, and it converges to the stable opinion vector given by:
\begin{equation}
   \bm{z} = (I-\Lambda W)^{-1} (I-\Lambda) \bm{u}\label{eq:fj_solution},
\end{equation}
where  $I$ is the identity matrix.
\end{theorem}
\noindent The final opinions computed via Theorem~\ref{theo:fj_convergence} serve as the baseline in Section~\ref{sec:performance_evaluation} for comparison with the outcomes of the Friedkin–Johnsen model on cascades.

This version of the FJ model is \emph{synchronous}, meaning that all nodes update their opinion simultaneously. However, in reality, people engage in asynchronous interactions, so opinions do not evolve simultaneously. In other words, we do not engage with all our social contacts all at once. 
%
%
To capture this, the FJ model also has an asynchronous version~\cite{frasca2013gossips} that, as we will discuss later, has been proven to converge to the synchronous FJ model of Equation~\eqref{eq:fj}. In the asynchronous model,  an edge $(i,j)$ is selected uniformly at random at each time step and the opinion of \emph{only} one of the two nodes (say $i$, selected at random) is updated, according to the following equation
\begin{equation}
     x_i(t+1) = h_i \left[(1-\gamma_{ij})x_i(t)+ \gamma_{ij}x_j(t)\right]+(1-h_i)u_i,
     \label{eq:fj_asyn_update_i}
\end{equation}
\noindent while the rest of the nodes maintain their opinion.  Note that the term $(1 - \gamma_{ij})x_i(t)$ balances the current expressed opinion of node $i$ against that of its neighbor $j$, with $\gamma_{ij}$ controlling the strength of influence from $j$.
Here matrices $H = \{h_{ij}\}$ and $\Gamma = \{\gamma_{ij}\}$ replace $\Lambda$ and $W$ of the synchronous model, and are different from them. 
In fact, to guarantee the convergence of the model in Equation~\eqref{eq:fj_asyn_update_i} to that in Equation~\eqref{eq:fj}, the formulation has to take into account the probability of node $i$ being randomly selected, which, in turn, depends on the degree of nodes. 
To guarantee convergence to the synchronous FJ model, $H$ and $\Gamma$ should be defined from $\Lambda$ and $W$ as follows~\cite{frasca2013gossips}:
    \begin{align}
        &h_i =1-\frac{1-\lambda_i}{d_i}\label{eq:h_i}\\
        &\gamma_{ij} = \left\{\begin{array}{ll}
             \frac{d_i(1-h_i)+h_i-(1-\lambda_iw_{ii})}{h_i}& \textrm{if }i=j\\
             \frac{\lambda_i w_ij}{h_i} & \textrm{if }i\neq j
        \end{array}\right.\label{eq:gamma_ij}
    \end{align}
where $d_i$ is the degree of node $i$. 
In~\cite{frasca2013gossips}, Frasca~\etaltwo establishes the ergodic convergence of the asynchronous FJ algorithm defined by Equation~\eqref{eq:fj_asyn_update_i} to the classical FJ. Notably, while the individual state variables generally do not converge in the pointwise sense and often exhibit persistent oscillation, their time-averaged values converge asymptotically to a stable limit, which, in our case, coincides with the stable point of the FJ. The FJC model is developed by building upon the asynchronous version of the FJ dynamics, as detailed in Section~\ref{sec:fjc}, which allows us to model pairwise opinion influence and embed the FJ mechanism within the cascade process. 

\vspace{-8pt}
\subsection{Polarization}
\label{sec:polarization_preliminaries} 

In~\cite{biondi2023dynamics}, the most widely used polarization metrics in the related literature are discussed and classified according to their underlying semantics.
We summarize below a selected subset of these definitions, which are used in Section~\ref{sec:performance_evaluation}.
%
\begin{definition}\label{def:polarization indices} For an opinion $\bm{x}=(x_i)\in [-1,1]^v$ the following polarization indices are defined:
\begin{align}
P_{2}(\bm{x}) &= \frac{1}{|\mathcal{V}|} \sum_{i \in \mathcal{V}} x_i ^2 = \frac{1}{|\mathcal{V}|} \| \bm{x} \|_2^2 \label{eq:p2}\\
P_{3}(\bm{x}) &=  \sum_{i \in \mathcal{V}} x_i ^2 = \| \bm{x} \|_2^2 \label{eq:p3}\\
P_4(\bm{x}) &=  \sum_{i \in \mathcal{V}} |x_i| = \| \bm{x} \|_1 . \label{eq:p4}
\end{align}
\end{definition}
In the context of measuring polarization in social networks, these metrics consider polarization to occur when opinions are significantly distant from a neutral opinion, which is assigned a value of 0. 

A dynamic opinion model is considered \emph{polarizing} with respect to a given metric $\Phi$ if there exists at least one initial opinion vector $\bm{u}$ such that the corresponding final opinion  $\bm{z}$ exhibits greater polarization, i.e., $\Phi(\bm{z}) > \Phi(\bm{u})$. The quantity
\begin{equation}
    \Delta_\Phi(\bm{u}) = \Phi(\bm{z})-\Phi(\bm{u}) \label{eq:polarization}
\end{equation}
is referred to as the \emph{polarization value} and, by definition, is positive if and only if the model is polarizing with respect to the metric $\Phi$.
Our previous research~\cite{biondi2023dynamics} showed that the FJ model is inherently polarizing and provided formulas to derive polarizing vectors for the polarization metrics descibed above. These vectors are closely linked to the spectral properties of the matrix $H = (I - \Lambda W)^{-1}(I - \Lambda)$, which maps initial opinions $\bm{u}$ to final opinions $\bm{z}$ as in Equation~\eqref{eq:fj_solution}. In this paper, we use the following polarizing initial conditions:
\begin{itemize}
\item $\bm{u}_{B_1(1)}$: maximizes polarization under $P_4$ within the $L_1$-ball of radius 1;
\item $\bm{u}_{B_2(1)}$ and $\bm{u}_{B_2(t)}$: maximize polarization under $P_2$ and $P_3$ within the $L_2$-balls of radius 1 and $t$, respectively;
\item $\bm{u}_{V_{>1}}^{heu}$: a heuristic approximation that maximizes $P_2$ and $P_3$ over a larger subdomain of the opinion space.
\end{itemize}
Analysis of these vectors in~\cite{biondi2023dynamics} revealed which nodes contribute most to polarization shifts in the network. Polarization is maximized when extreme opinions are assigned to the most influential individuals in the network, while less influential or more susceptible individuals are initialized with neutral opinions. Accordingly, these vectors place opinions close to the boundaries of the opinion domain (i.e., near $\pm1$) on nodes that are either highly PageRank-central or highly stubborn, while assigning opinions close to zero to nodes that are more susceptible and/or less PageRank-central. The analysis in~\cite{biondi2023dynamics} confirmed that PageRank centrality is the centrality measure that better captures the influence among nodes in social networks~\cite{proskurnikov2017tutorial}.

In Section~\ref{sec:performance_evaluation}, we focus on these polarizing initial conditions to study the effect of information cascades under  \emph{worst-case} polarization scenarios for the baseline FJ model. While alternative initial opinion profiles would generally lead to different quantitative outcomes, these configurations are known to maximize polarization under FJ dynamics. While they are not guaranteed to play the same role under the FJC dynamics, they provide a principled reference point to assess how the introduction of information cascades modifies polarization. As we will show, information cascades tend to amplify the influence of already central individuals, who are precisely emphasized by the polarizing configurations considered here. For this reason, these scenarios provide a principled and informative reference point to assess how cascades can amplify or mitigate polarization.

\vspace{-8pt}

\section{The Friedkin-Johnsen on Cascades model}
\label{sec:model}


In this section, we introduce the Friedkin–Johnsen on Cascades (FJC) model, which combines opinion updating with cascade-based information diffusion. The FJ model describes opinion formation through interpersonal influence, while information cascades describe the propagation of content in online networks. The FJC model explicitly integrates these mechanisms, capturing how opinions are updated in response to exposure to posts in a cascade. This framework provides a basis for studying how information diffusion interacts with opinion dynamics in networked environments.
Throughout this section, we introduce several modeling assumptions, which are discussed where they become relevant in order to facilitate understanding of the proposed framework. For the reader’s convenience, all assumptions are summarized in Table~\ref{tab:assumptions}.

\begin{table*}
\centering
\caption{Summary of modeling assumptions in the FJC framework}
\label{tab:assumptions}
\setlength{\tabcolsep}{6pt}
\begin{tabularx}{\textwidth}{c l X}
\toprule
\textbf{ID} & \textbf{Assumption} & \textbf{Description} \\
\midrule
A1 & Opinion-neutral content  
& Shared content does not carry an intrinsic opinion; opinions are conveyed only by the individuals sharing it. \\
A2 & Tree-structured exposure 
& A post propagates from the seed in a tree-like manner across successive neighbour layers, with each node influenced only upon first exposure. \\
A3 & Sequential cascade propagation 
& Only one post propagates at a time; simultaneous cascades are not considered. \\
\bottomrule
\end{tabularx}
\vspace{-10pt}
\end{table*}

\vspace{-10pt}
\subsection{Integrating Independent Cascades into the Friedkin–Johnsen Framework}\label{sec:model_setting}

\begin{figure}[t]
\centering
    \includegraphics[width=0.4\textwidth]{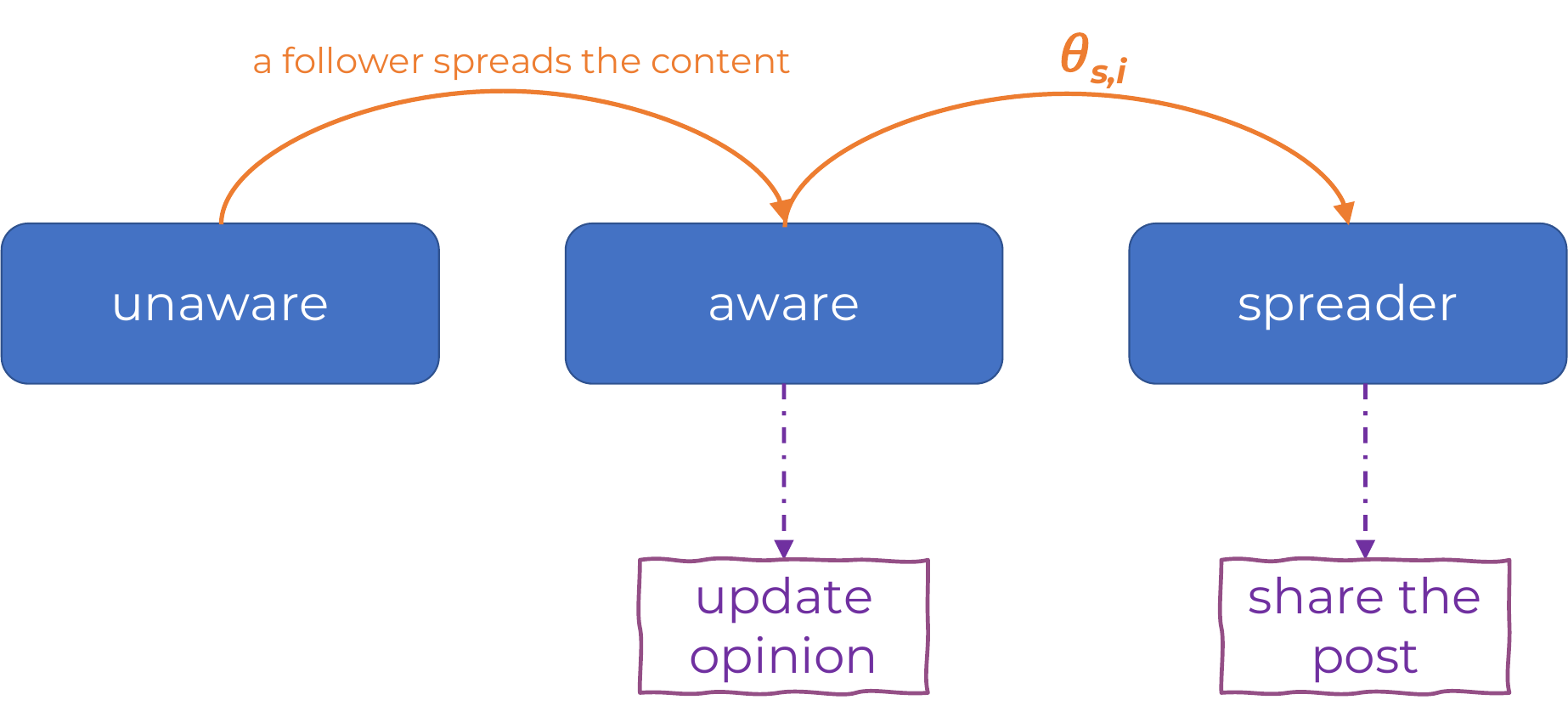}
    \caption{States, transitions and actions of the FJC model for a single node. The blue boxes indicate the states and the orange arrows with corresponding probabilities indicate the transitions between states. The purple arrows and boxes indicate the actions made within the corresponding states.}
        \label{fig:actions_states}
    \vspace{-10pt}
\end{figure}

The standard IC model is adapted here to integrate opinion updating into the cascade process, yielding what we refer to as the FJC framework. Rather than restating the full IC mechanism, we explicitly highlight the key differences with respect to the classical model. 
First, nodes transition through three states—\emph{unaware}, \emph{aware}, and \emph{spreader}. Becoming aware of a post triggers an opinion update, a feature absent in the standard IC model (see Figure~\ref{fig:actions_states}). 
Second, information propagation is constrained to a directed tree structure rooted at the source of the post. Nodes are organized into successive neighbourhoods of increasing distance from the source, which we refer to as \emph{layers}. Under this assumption, each node updates its opinion only upon its first exposure to the post, that is, when it is reached by resharing neighbours in the immediately preceding layer. During this exposure event, a node may aggregate influence from all such neighbours, while subsequent resharings do not induce further opinion updates. 
As a consequence, the cascade unfolds over a directed tree in which each node appears in at most one layer and is influenced only by its predecessors.

We assume (A1) that the content being shared does not express an opinion on its own, but rather the opinion is conveyed only by the person sharing it along with a comment. This assumption allows us to avoid considering the post as an external factor that can modify opinions and instead to focus on studying how the simple cascade dynamic affects opinion formation in comparison to the FJ scenario.

In our model, an individual identified as the source disseminates a piece of news to the online social network. In the rest of this paper, we streamline our discussion by initially constructing the model under the assumption that each post originates from a single seed node. However, it is worth noting that the model's flexibility allows for straightforward extension to accommodate multiple sources, akin to the original information cascade model. In fact, including multiple sources simply affects the first level of the information cascade tree, while the rest remains unchanged. 
We assume (A2) that once the post is shared the first time by the seed node $r$, it spreads outwards like a tree, from the original source to the first, second, third, and subsequent follower neighbours (i.e. in-neigbours according to notation of Section~\ref{sec:social_network}), which we will refer to as the nodes in the first, second, third, ... \emph{layer}. Under this assumption, each node can appear in at most one layer of the cascade tree, corresponding to its first exposure to the post. While a node may be influenced by multiple predecessors in the previous layer, all such influences occur simultaneously at this first exposure. 
Following this initial encounter, we maintain that nodes typically become disinterested in posts shared by others. Thus, we refer to the $l$-th in-neighbourhood as the $l$-th \emph{layer} (or \emph{layer} $l$), and denote it by $\mathcal{O}_l^{(r)}$.
The depth of the graph is given by the eccentricity of the root node, denoted by $\epsilon^{(r)}$, which corresponds to the maximum distance from $r$ to any other node.
Accordingly, the entire tree comprising all nodes in the \emph{aware} state of the cascade is represented as the sequence $\mathcal{O}^{(r)} = \left(\mathcal{O}l^{(r)}\right)_{l=0}^{\epsilon^{(r)}}$ of all layers. Please note that, unlike the standard IC model, where influence attempts may occur independently along multiple incoming edges, here the cascade is explicitly represented as a directed tree rooted at the source, ensuring a unique first exposure for each node.

\begin{figure}
    \centering
        \begin{subfigure}[b]{0.20\textwidth}
      \centering
         \includegraphics[width=0.65\textwidth]{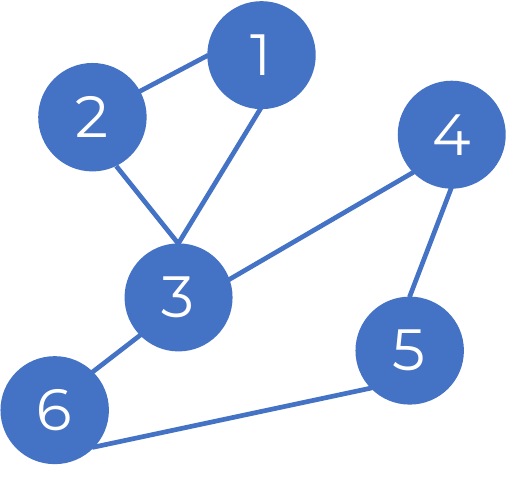}
         \caption{Network}
     \end{subfigure}
     \begin{subfigure}[b]{0.20\textwidth}
     \centering
         \includegraphics[width=0.9\textwidth]{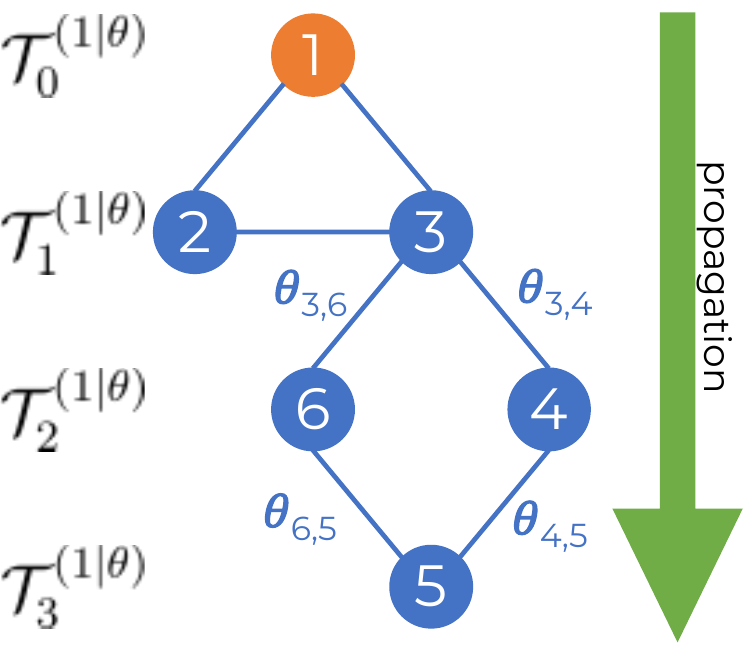}
     \caption{$\mathcal{T}^{(1|\theta)}$} 
    \end{subfigure}
       \caption{A toy example of a network in (a) and the corresponding propagation tree for $r=1$ in (b). Nodes $2$ and $3$ update their opinion deterministically, while the others do it with probability $\theta$.}
        \label{fig:example}
\end{figure}

A  post $p$ generated at node $r$ spreads into the network over the tree $\mathcal{O}^{(r)}$. However, not all nodes in this tree receive the content, because there is a probability $\theta$ that each node will not reshare the post.  The subtree of  $\mathcal{O}^{(r)}$ that includes only nodes that have received the post is denoted as $\mathcal{T}^{(r,\theta)} = (\mathcal{T}^{(r,\theta)}_{l} )_{l=0,\dots,\epsilon^{(r)}}$, and it holds that $\mathcal{T}^{(r,\theta)}_{l}\subseteq \mathcal{O}^{(r)}_{l}$ for each $l$. $\mathcal{T}^{(r,\theta)}$ is  a stochastic realization of the tree $\mathcal{O}^{(r)}$, as indicated by the superscript~$\theta$. In Figure~\ref{fig:example}, there is a simple example of a network and a corresponding propagation tree. 
We can describe it in formal terms in the following way. We refer to the \emph{cascade} generated by the post $p$ originating in node $r$ as the collection of nodes in the tree $\mathcal{T}^{(r,\theta)}$ and we will denote it with $\mathcal{C}^{(r)}$. Please observe that nodes in layer~1 always belong to the cascade because they receive the post directly from the source, while those of layer $l\geq2$ belong to the cascade if and only if they receive the post from nodes in layer $l-1$.\\
Let us indicate with $S_i$ the random variable modelling the event of resharing a post for a node $i\neq r$ (it takes the value~1 when the post is reshared, 0 otherwise). 
%
%
Then, the tree $\mathcal{T}^{(r,\theta)}$ is described as:
\begin{equation}
    \left\{
    \begin{array}{l}
        \mathcal{T}^{(r,\theta)}_{0} =\{r\} \\
        \mathcal{T}^{(r,\theta)}_{1} =\mathcal{N}^-(r) \\
        \mathcal{T}^{(r,\theta)}_{l+1} = \left(\bigcup\limits_{\{i\in \mathcal{T}^{(r,\theta)}_l:S_i=1\}} \mathcal{N}^-(i)\right)\setminus \mathcal{T}^{(r,\theta))}_{l}.
    \end{array}
    \right.   
\end{equation}
Please, remember that $\mathcal{N}^-(i)$ refers to the in-neighbours of node $i$. This construction intentionally departs from the classical IC formulation to isolate the effect of single-exposure, cascade-driven opinion updates, which is essential for coupling the cascade process with the Friedkin-Johnsen dynamics.\\
For our convenience, for each node $i\in\mathcal{O}^{(r)}_{l}$ we denote with $\varphi(i)$ the set of nodes who actually influence $i$'s opinion in the stochastic realization of the cascade, i.e the nodes in layer $l-1$ that are in $i$'s out-neighbour and reshare the post. In other terms:
\begin{equation}\label{eq:predecessors}
    \varphi(i)=\left\{j\in \mathcal{N}^-(i): j\in\mathcal{T}^{(r,\theta)}_{l-1} \wedge S_{j}=1  \right\}
\end{equation}
We will refer to this set as the \emph{predecessors} of $i$.

\vspace{-8pt}
\subsection{The Cascade-Driven Friedkin-Johnsen}
\label{sec:fjc} 

We will now discuss how to integrate the independent cascade model described in Section~\ref{sec:model_setting} with the FJ model from Section~\ref{sec:fj}, resulting in a FJ variant (FJC) where opinions are influenced by the sharing of posts. 
The FJC model operates iteratively, following specific states and transitions for each node (see Figure~\ref{fig:actions_states}). At the start of each cascade, every node, except for the designated seed node $r$, begins in the unaware state. During each iteration, nodes $i$ that are exposed to a post from another node $s$ undergo two consecutive actions. 
First, these exposed nodes transition to the "aware" state and update their opinions regarding the post. Following this, with a probability $\theta_{s,i}$, these nodes enter the spreader state, sharing the post again if they belong to a layer $l$ that is smaller than the maximum layer $\epsilon^{(r)}$. If a node does not meet this condition, the propagation of the post stops, and a new cascade is initiated.
We will examine the spreading process of $N$ posts, denoted with $p_1,\dots,p_N$. We make an assumption (A3) that posts propagate one at a time, meaning that simultaneous spread of two posts is not possible. While this assumption is stringent, it aids in designing the model .
In future work, we will explore ways to relax this assumption and understand its impact on the model. In Section~\ref{sec:eval_validation}, we discuss the model effectiveness in real-world settings.

Now, we want to discuss the details of the opinion updating operation. Please observe that, since we want to update opinion layer by layer, we need an asynchronous version of the FJ, similar to that of Equation~\eqref{eq:fj_asyn_update_i}.  
We focus on the spreading process of the $n$-th post $p_n$ generated at node $r_n$, which involves nodes in the cascade $\mathcal{C}^{(r_n)}$. As the cascade starts, the opinions of nodes are equal to $\bm{u}=\bm{x}(0)$ if $n=1$, i.e. if we are considering the spreading of the first post. Otherwise, they are equal to the final opinion of nodes as the previous post $p_{n-1}$ completed spreading. 
Every time a node $j$ in the cascade reshares the post $p_n$ 
at a time $t$, it conveys its current opinion $x_j(t)$ together with the reposting.  A node $i$ in layer $l$ can receive the post (which, as explained above, is a proxy for a node's opinion) from one or more predecessors $j$. This triggers the update of node $i$'s opinion, which is carried out according to the following update rule:

{\small
\begin{align}
    x_{i}&(t)
    = \nonumber\\
    &=\frac{1}{|\varphi(i)|}
    \Bigg\{
    \sum_{j\in\varphi(i)} h_{i}
    \Big[(1-\gamma_{ij})x_{i}(t-1)
    + \gamma_{ij}x_{j}(t-1)\Big] +\nonumber\\
    &\qquad\qquad+ (1-h_{i})u_{i}
    \Bigg\}
    \label{eq:fjc_update}
    \vspace{-5pt}
\end{align}
}
%
\noindent
where $h_i$ and $\gamma_i$ are defined as in Equations~\eqref{eq:h_i} and~\eqref{eq:gamma_ij}, and $\varphi(i)$ is defined in the Equation~\eqref{eq:predecessors}.
The above update rule factors in the average total influence on node~$i$ by all its in-neighbours that have reshared the post in the previous layer. This means that the average is calculated only from those neighbors who have already engaged with the post and shared their opinions, thereby influencing the opinion of node $i$. Note that, differently from the asynchronous FJ update in Equation~\eqref{eq:fj_asyn_update_i}, node $i$ assesses the opinions of its predecessors all at once, not in a pairwise fashion. 

Differently from the FJ model in Equation~\eqref{eq:fj}, we update opinions node by node, following the order provided by the layer of tree $\mathcal{T}^{(r_n,\theta)}$. 
Once the opinions of a single layer are updated, the post spreads again. 
In Algorithm~\ref{alg:fjc} we summarize the entire FJC process.

\begin{algorithm}
\caption{FJC}\label{alg:fjc}
\begin{algorithmic}[1]
\Require{$\bm{R}$ (vector of  post seeds), $\bm{u}$ (initial opinion vector)} 
\Ensure{$\bm{x}$ (final opinion vector)}
\State $\bm{x} \gets \bm{u}$
\Comment{initialize the opinion vector}
\For{$r$ in $\bm{R}$}
    \For{$l$ in $\{1,\dots,\epsilon^{(r)}\}$} \Comment{iterate over layers}
            \For{$i$ in $\mathcal{T}^{(r,\theta)}_l$}
                \State $x_{i} \gets$ Equation ~\eqref{eq:fjc_update} \Comment{update opinions}
            \EndFor
    \EndFor
\EndFor
\end{algorithmic}
\end{algorithm}

\vspace{-15 pt}

\section{Performance Evaluation}
\label{sec:performance_evaluation}

In this section, we analyze the performance and validate the FJC model across different social networks. To this end, we conducted two distinct sets of experiments. 
The first set of experiments (Sec.~\ref{sec:eval_validation}) focused on validating the FJC model using real-world cascade data. Here, we evaluated how well the model can approximate opinion evolution based on traces from an online social network, where assumptions (A2) and (A3) do not always hold. Specifically, (A2) posits that cascades spread in a tree-like fashion from the seed node, layer by layer, and (A3) assumes that only one post propagates at a time. Assumption (A1), however, could not be relaxed, as it underpins the model’s formulation of opinion dynamics  — it assumes that the content itself is opinion-neutral, and opinions are conveyed solely by the user sharing the post. In the following, we provide a detailed description of the experimental settings and discuss the results obtained.
The second set (Sec.~\ref{sec:eval_synth_casc}) aimed to compare the opinion dynamics produced by the FJC model with those of the classical FJ model. For this purpose, we used three social networks—two real-world and one synthetic—and generated synthetic cascades on top of them. Specifically, we selected 15 seed nodes per graph to initiate the simulated cascades.

\vspace{-8pt}
\subsection{Common Experimental Settings} 

All networks used are unweighted, so the social matrix $\hat{W}$ corresponds to the adjacency matrix. As described in Section~\ref{sec:social_network}, the influence matrix is derived by row-normalizing $\hat{W}$: $w_{ij} = \frac{\hat{w}{ij}}{\sum_k \hat{w}{ik}}$. Following our previous work~\cite{biondi2023dynamics}, we considered three strategies for assigning node susceptibility. In the first two, susceptibility is coupled to PageRank centrality, which is known to be strongly correlated with node influence in linear opinion dynamics, including the Friedkin–Johnsen model~\cite{proskurnikov2017tutorial}. While PageRank does not exactly coincide with the influence weights induced by FJ dynamics, it provides a well-established proxy for identifying influential nodes in directed networks. 
Specifically, for each node $i$ with PageRank score $\mathrm{P}_i$, we assign susceptibility $\lambda_i$ either proportional to $\mathrm{P}_i$ or to $\mathrm{P}_i^{-1}$, with values rescaled to the interval $(0,1)$. These two configurations allow us to study \emph{stressed} scenarios in which susceptibility is systematically aligned or anti-aligned with node influence.  To assess the role of this coupling, we also consider a control setting in which all nodes are assigned the same susceptibility value, $\lambda_i = 0.6$, thereby fully decoupling susceptibility from centrality. 
Finally, since our primary goal is to study the impact of cascades on polarization, we initialized opinions using the polarizing vectors defined in Section~\ref{sec:polarization_preliminaries}, ensuring a consistent and controlled setup for measuring polarization dynamics. Final opinions under cascades were then compared with those of the FJ model, whose outcomes are analytically derived via Equation~\eqref{eq:fj_solution} using the same parameters, except for the reposting probability~$\theta$, which is not applicable to FJ.

\vspace{-8pt}
\subsection{FJC with Real Cascades}
\label{sec:eval_validation}

\subsubsection{Dataset description and experimental setting}
We validate the model using a real-world dataset from Bhowmick~\etaltwo~\cite{bhowmick2019temporal}, which contains Twitter (now X) data related to the 2015 Nepal earthquake. This dataset is well-suited for our analysis as it includes both the follower network and tweet/retweet histories. Unlike the synthetic networks in Sec.~\ref{sec:eval_synth_casc}, this is a directed graph, with multiple connected components. Since FJ-based models operate within a connected structure, we extract the largest strongly connected component (LSCC). Some cascades span disconnected components, suggesting hidden links, so we restrict our analysis to cascades occurring entirely within the LSCC. The resulting graph contains 220,974 nodes, 16,772,693 edges, and 32,313 cascades.
%
Initial opinions are set using the polarizing vectors described in Section~\ref{sec:polarization_preliminaries}. However, their computation requires computing the eigensystem of an $n\times n$ dense matrix, which becomes impractical due to the extensive size of the network, even when employing optimized algorithms. To tackle this challenge, we derived a subgraph known as \textit{Nepal top cascades}, which includes nodes from the 500 longest cascading events. The rationale behind focusing on the longest cascades is that they inherently present a greater challenge for our model, as they involve more complex and prolonged interactions among nodes and increase the likelihood that cascades can overlap (thus, violating assumption (A3)). Specifically, we ensured that all nodes participating in these cascades are included in the subgraph using a heuristic for the directed Steiner Tree problem~\cite{watel2016practical}. Table~\ref{tab:Nepal_datasets} summarizes the key properties of this subgraph, which encompasses over 6,000 cascades. To assess the robustness of our results with respect to the selection of the top-cascade subgraph, we replicated the full experimental analysis on an additional sampled subgraph; the corresponding results, which are qualitatively consistent with those reported here, are presented in the SI.

\begin{table}[t]
\centering
\caption{Summary of the  real-world Nepal dataset subgraphs used for validation}
\label{tab:Nepal_datasets}
\setlength{\tabcolsep}{2.5pt}
\begin{tabular}{ ccccc }
\toprule
  &\textbf{\#nodes}& \textbf{\#edges} &  \textbf{\#cascades} & \textbf{max cascade length}\\
 \midrule
 Nepal top cascades & 26321  & 4222064  & 6083 &  21\\
\bottomrule
\end{tabular}\vspace{-7pt}
\end{table}

The dataset $D$ provides a sequence of tuples $\mathcal{C}_D = \big((t_1, p_1, i_1), (t_2, p_2, i_2), \dots\big)$, where node $i_k$ shares post $p_k$ at time $t_k$, representing the temporal evolution of cascades. We assume that this sequence of reshares drives the opinion dynamics in the real world: each reshare is treated as an expression of opinion. At time $t_k$, when node $i_k$ shares a post, its followers update their opinions according to Equation~\eqref{eq:fjc_update}. This process relies solely on assumption (A1); assumptions (A2) and (A3) are not used. We refer to this approach as \emph{real opinion dynamics}. It is a data-driven heuristic and cannot be analyzed analytically.

Our goal is to evaluate how well this real dynamic is captured by the proposed FJC model (we also include the classical FJ model for reference). For the FJC model, we need to estimate the resharing probability $\theta_i$ for each node. Inspired by weighted cascade models~\cite{kempe2003maximizing,tang2015influence}, we define $\theta_i$ as the ratio between the number of posts reshared by node $i$ and the number of posts it has seen from its followees. Table~\ref{tab:theta} shows the distribution of $\theta_i$, which is generally low, indicating that most nodes rarely reshare posts. Please note that these values are consistent with the retweet probabilities in~\cite{Romero_Meeder_Kleinberg_2011}, remaining below 0.02.
Unlike the \emph{real} dynamics, the FJC model processes cascades in a tree manner and sequentially. Specifically, posts are propagated from the sources to the network \emph{layer-by-layer}, in accordance with assumption (A1) and each cascade starts only after the previous one ends, in accordance with assumption (A3).

\begin{table}
\centering
\caption{Statistics of followers, followed and $\theta$ in the Nepal subgraph}\label{tab:theta}
\setlength{\tabcolsep}{2.5pt}
\begin{tabular}{ cccccccc }
\toprule
\multicolumn{8}{c}{\textbf{Nepal top cascades}}\\
 & Min. &1st Qu.&  Median & 3rd Qu.&    Max. &  Mean &    N.A.\\
 \midrule
\textbf{Followers} & 1.0    & 23.0   & 59.0  & 148.0 & 11022.0 & 160.4 &-\\
\textbf{Followees} & 1.0  &  53.0  & 109.0  & 204.0 & 6851.0 & 160.4 &-\\
$\theta$ & 0.0000 & 0.0000 &  0.0000 & 0.0000 & 1.0000 & 0.0156 & 681\\
\bottomrule
\end{tabular} \vspace{-10pt}
\end{table}
%

\subsubsection{Opinion evolution of real opinion dynamics vs FJC and FJ} 
\begin{figure}
\centering
        \includegraphics[width=0.5\textwidth]{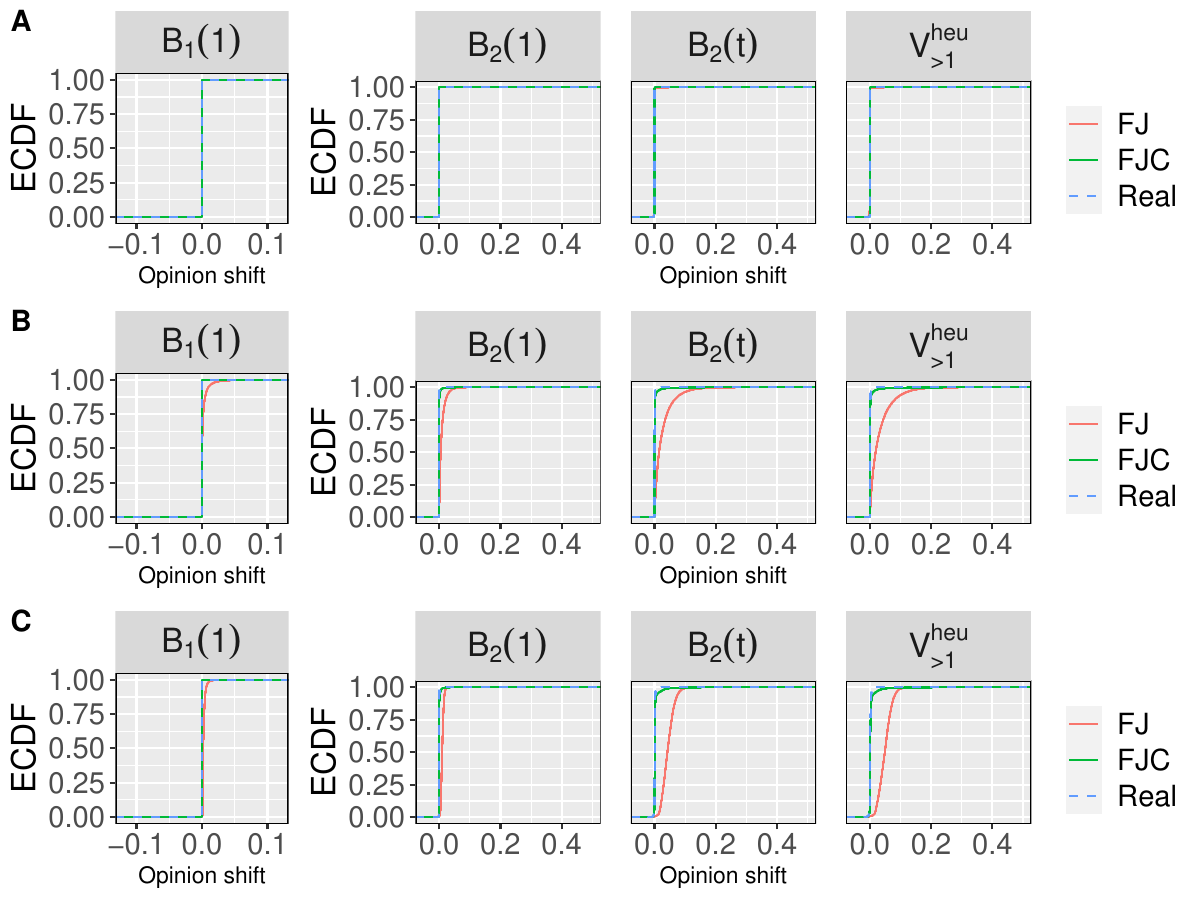}
 \caption{ECDF of the opinion shift of nodes, i.e. final opinion minus initial opinion, for the Nepal top cascades in the three different opinion dynamics. (A) $\lambda_i \sim \mathrm{P}_i$, (B) $\lambda_i \sim \mathrm{P}_i^{-1}$, (C) $\lambda_i = 0.6$. }
 \label{fig:opinion_shift}
 \vspace{-10pt}
\end{figure}

Due to the large size of the networks, we cannot visualize the polarizing vectors, 
since the plots would be unreadable. Instead, our main objective with these datasets is to evaluate the robustness of the FJC model against the \emph{real} opinion dynamics. We do so by comparing the distributions of opinion shifts—i.e., the differences between final and initial opinions—for the real evolution (labeled \textit{Real}) and the predictions of FJC and FJ. As shown in Figure~\ref{fig:opinion_shift}, FJC closely mirrors the real dynamics observed in the \emph{Nepal top cascades} dataset, with the FJC and \textit{Real} curves largely overlapping, outperforming the standard FJ model. 
Please note that long cascades are more likely to violate assumption (A3), reducing FJC’s accuracy. Nevertheless, FJC remains surprisingly reliable even in this more challenging scenario. 


\subsubsection{Polarization} 
\begin{figure}
\centering
    \includegraphics[width=0.5\textwidth]{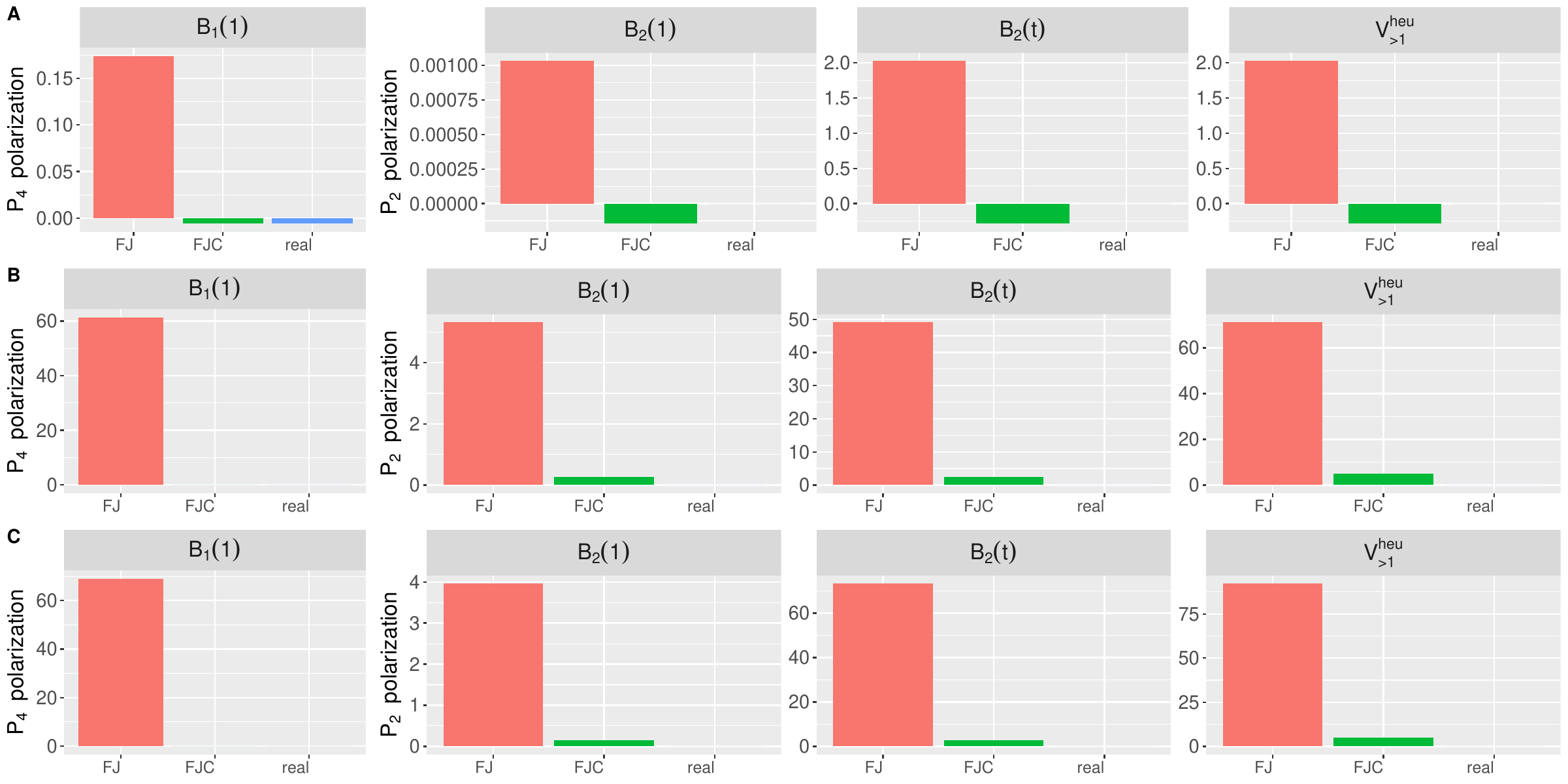}
    \caption{Polarization value of the different vectors in the corresponding metrics in the three different opinion dynamics for Nepal top cascades. (A) $\lambda_i \sim \mathrm{P}_i$, (B) $\lambda_i \sim \mathrm{P}_i^{-1}$, (C) $\lambda_i = 0.6$. }
        \label{fig:nepal_pol}
        \vspace{-10pt}
\end{figure}

Figure~\ref{fig:nepal_pol} illustrates the polarization outcomes across different settings. The \emph{Real} opinion dynamics remain almost entirely null, since opinion shifts are negligible in all scenarios (Figure~\ref{fig:opinion_shift}). Similarly, the cascade-based dynamics (FJC) exhibit very low—or even slightly negative—polarization. Both stand in sharp contrast to the classical FJ model, which consistently produces positive polarization with substantially higher magnitude. This behavior can be explained by Table~\ref{tab:theta}, which shows that most nodes have $\theta$ values close to zero, indicating very limited reposting activity. Consequently, the opinion dynamics remain largely inactive, preventing polarization from emerging. In Section~\ref{sec:eval_synth_casc}, we will further validate this hypothesis on synthetic networks, also exploring scenarios with larger $\theta$ values to assess their impact on polarization.
%

\vspace{-8pt}

\subsection{FJC vs FJ --  with Synthetic Cascades}
\label{sec:eval_synth_casc}

\subsubsection{Dataset description and experimental setting}
In this section, we evaluate the performance of the FJC model through simulations on three social networks: the Karate Club graph~\cite{zachary1977information}, a Facebook graph~\cite{leskovec2012learning}, and a synthetic Barabási-Albert network~\cite{barabasi2016network}. The Karate Club graph is a well-known small network with 34 nodes, while the Facebook graph includes 1519 nodes from the largest connected component of a 4039-node snapshot. The Barabasi-Albert network, which captures the scale-free structure of real social networks, was generated with 100 nodes and parameter $k=6$, meaning each new node connects to 6 existing nodes. Please refer to the SI for a graphic visualization of the graphs. 
Since cascades are simulated, we randomly selected 15 seed nodes per network to initiate the posts and kept them fixed across all simulations, in order to control for variability due to source selection. The properties of the selected seed nodes and their centrality profiles are analyzed in the SI, where we show that this choice does not introduce systematic bias. We set the resharing probability $\theta_{s,i}$ uniformly to a single value $\theta$ for all nodes, and tested three different scenarios. In the first diffusion scenario, we set $\theta = 0.01$, which is approximately the average value found in the Nepal dataset (see Table~\ref{tab:theta}). To explore other potential values, we referenced typical ranges established in the literature. Hodas et al. in~\cite{Hodas_Lerman_2012} estimated that the retweet probability for each tweet seen is around 0.15\% to 0.4\%.  
Based on this, we modelled a moderate diffusion scenario with $\theta = 0.1$, and for maximum diffusion, we used $\theta = 0.5$.
%
%
For each combination of network, susceptibility setting, reposting probability $\theta$, and initial opinion values, we performed 1000 independent simulation runs. 
%
%
\subsubsection{Opinion evolution under FJC vs FJ}

\begin{figure*}
\centering
\text{$\bm{\theta=0.01}$}\\[0.5em]  
        \includegraphics[width=\textwidth]{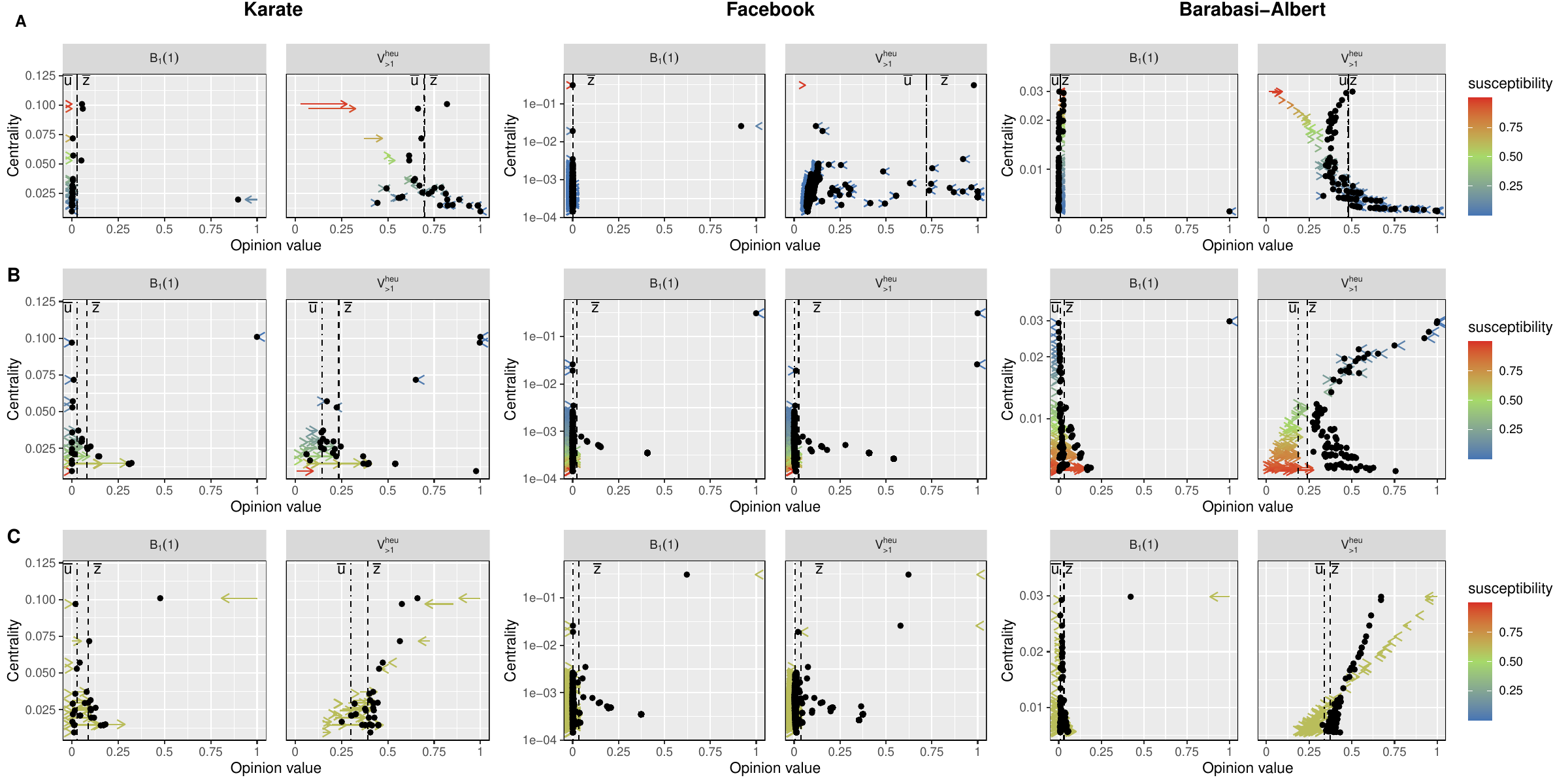}\vspace{-5pt}
 \caption{Polarizing opinions vs. node centrality with $\theta = 0.01$. Each arrow represents a node, starting at its initial opinion (prejudice) and ending at its final opinion under the FJC model. Arrow color indicates node susceptibility; black dots show the final opinions under the FJ model. (A) $\lambda_i \propto \mathrm{P}_i$; (B) $\lambda_i \propto \mathrm{P}_i^{-1}$; (C) $\lambda_i = 0.6$. Dot-dashed and dashed lines show the average initial and final FJC opinions.}
 \label{fig:results_001}
 \vspace{-10pt}
\end{figure*}

\begin{figure*}
\centering
\text{$\bm{\theta=0.1}$}\\[0.5em]  
        \includegraphics[width=\textwidth]{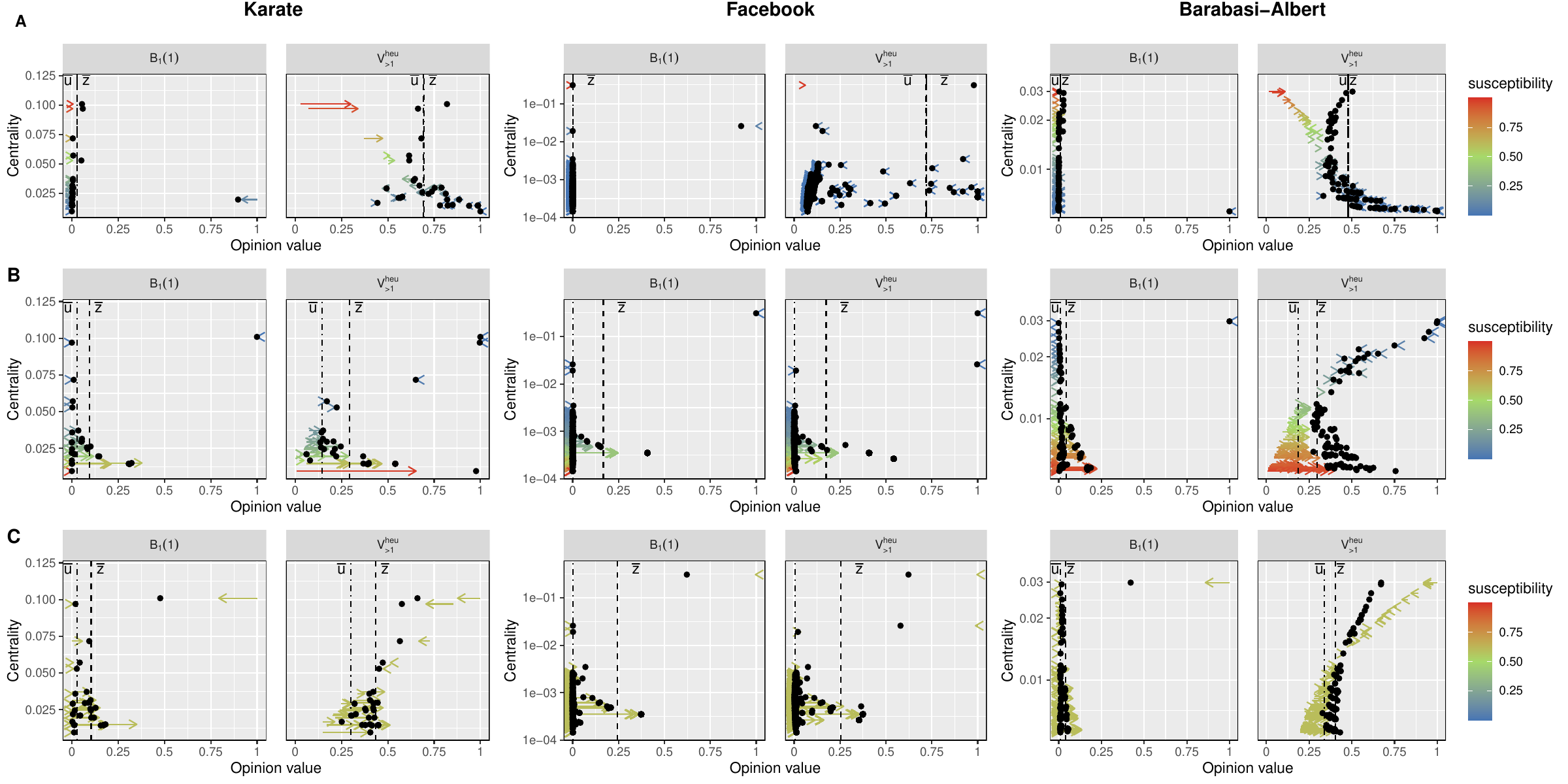}\vspace{-5pt}
 \caption{Polarizing opinions vs. node centrality with $\theta = 0.1$. Each arrow represents a node, starting at its initial opinion (prejudice) and ending at its final opinion under the FJC model. Arrow color indicates node susceptibility; black dots show the final opinions under the FJ model.  (A) $\lambda_i \propto \mathrm{P}_i$; (B) $\lambda_i \propto \mathrm{P}_i^{-1}$; (C) $\lambda_i = 0.6$. Dot-dashed and dashed lines show the average initial and final FJC opinions.}
 \label{fig:results_01}
 \vspace{-10pt}
\end{figure*}

\begin{figure*}
\centering
\text{$\bm{\theta=0.5}$}\\[0.5em]  
        \includegraphics[width=\textwidth]{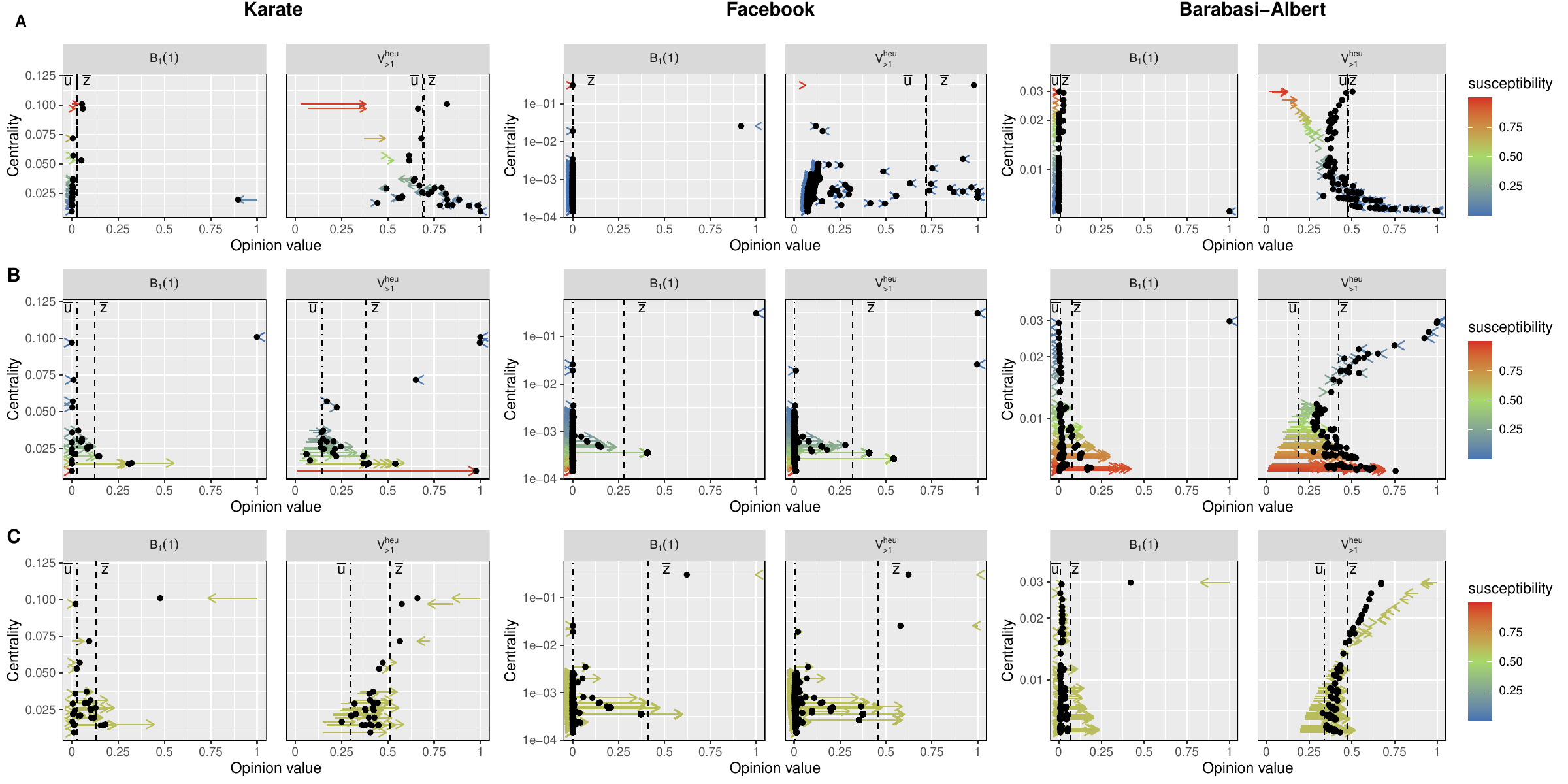}\vspace{-5pt}
 \caption{Polarizing opinions vs. node centrality with $\theta = 0.5$. Each arrow represents a node, starting at its initial opinion (prejudice) and ending at its final opinion under the FJC model. Arrow color indicates node susceptibility; black dots show the final opinions under the FJ model.  (A) $\lambda_i \propto \mathrm{P}_i$; (B) $\lambda_i \propto \mathrm{P}_i^{-1}$; (C) $\lambda_i = 0.6$. Dot-dashed and dashed lines show the average initial and final FJC opinions.}
 \label{fig:results_05}
 \vspace{-10pt}
\end{figure*}

We begin by analyzing how cascades influence final opinions under the FJC model compared to the classical FJ model. Due to space constraints, we will present $\bm{u}_{B_1(1)}$ and $\bm{u}_{V_{>1}}^{heu}$, which are 
the  local sub-optimal polarizing vectors that maximise $P_4$ and $P_2\text{-}P_3$ polarization, respectively, in the largest subdomain. Please refer to the SI for the full results. In Figures~\ref{fig:results_001}-\ref{fig:results_05}, each arrow represents a node’s opinion shift—from its initial opinion (prejudice) to its final FJC opinion. The black dot indicates the corresponding final opinion under the FJ model. Panels A and B show cases where susceptibility is proportional or inversely proportional to node centrality, respectively; panel C shows uniform susceptibility ($\lambda = 0.6$). 

We begin our discussion with Figure~\ref{fig:results_05}, which corresponds to the scenario with the highest propagation ($\theta = 0.5$), as this setting highlights the effects of cascades on opinion dynamics at a larger scale. Afterwards, we will examine how these trends change when $\theta$ is smaller.
Let us first consider panel C, where all nodes share the same susceptibility. Compared to the classical FJ model, highly central nodes in the FJC model tend to retain their initial opinions more strongly—evident from the shorter arrows relative to the distance to the FJ outcomes (black dots). This leads to a wider spread of final opinions among central nodes. In contrast, peripheral nodes are more easily influenced, and their opinions shift closer to those of the central ones. Overall, cascades amplify the role of network topology: central nodes act as opinion anchors, while peripheral nodes become more susceptible followers.
When susceptibility is inversely proportional to centrality (panel B), central nodes are already stubborn by design in both models. Final opinions are mostly similar across all networks, although we can observe slightly smaller opinion shifts in FJC. This reduction is due to the parameter $\theta$, which limits opinion propagation. This trend is further confirmed by the figures in the SI, where we also tested even larger values for this parameter: as $\theta$ increases, peripheral nodes shift more strongly in the FJC model, becoming increasingly influenced by central nodes.
Finally, with susceptibility proportional to centrality (panel A), network topology and susceptibility act in opposing directions. Peripheral nodes are nearly immobile due to low susceptibility, while central nodes—despite being more susceptible—change little due to their structural position. As a result, final opinions under FJC remain close to initial ones, more so than in the FJ model. This highlights how, in cascade-driven dynamics, centrality outweighs susceptibility, with central nodes maintaining influence regardless of the surrounding opinion mass.

Similar behaviors, though less pronounced, are observed with smaller values of $\theta$ ($\theta=0.01,0.1$) in Figures~\ref{fig:results_001} and~\ref{fig:results_01}.
In general, peripheral nodes exhibit reduced opinion shifts as $\theta$ decreases, reflecting their stronger dependence on the probability of content being reshared. Notably, in the Karate network, results remain almost unchanged across different $\theta$ values, as its small size allows most nodes to be reached by the cascade even when propagation is limited.
Interestingly, central nodes in the FJC model maintain stable behavior even with tighter propagation. This resilience is an intrinsic property of the model, stemming from the structured, sequential nature of its cascades. In fact, highly connected nodes are typically exposed to content early, when active sources are few, making them less susceptible to external influence.

\subsubsection{Polarization}
\label{sec:polarization}

\begin{figure}[h]
\centering
    \includegraphics[width=0.48\textwidth]{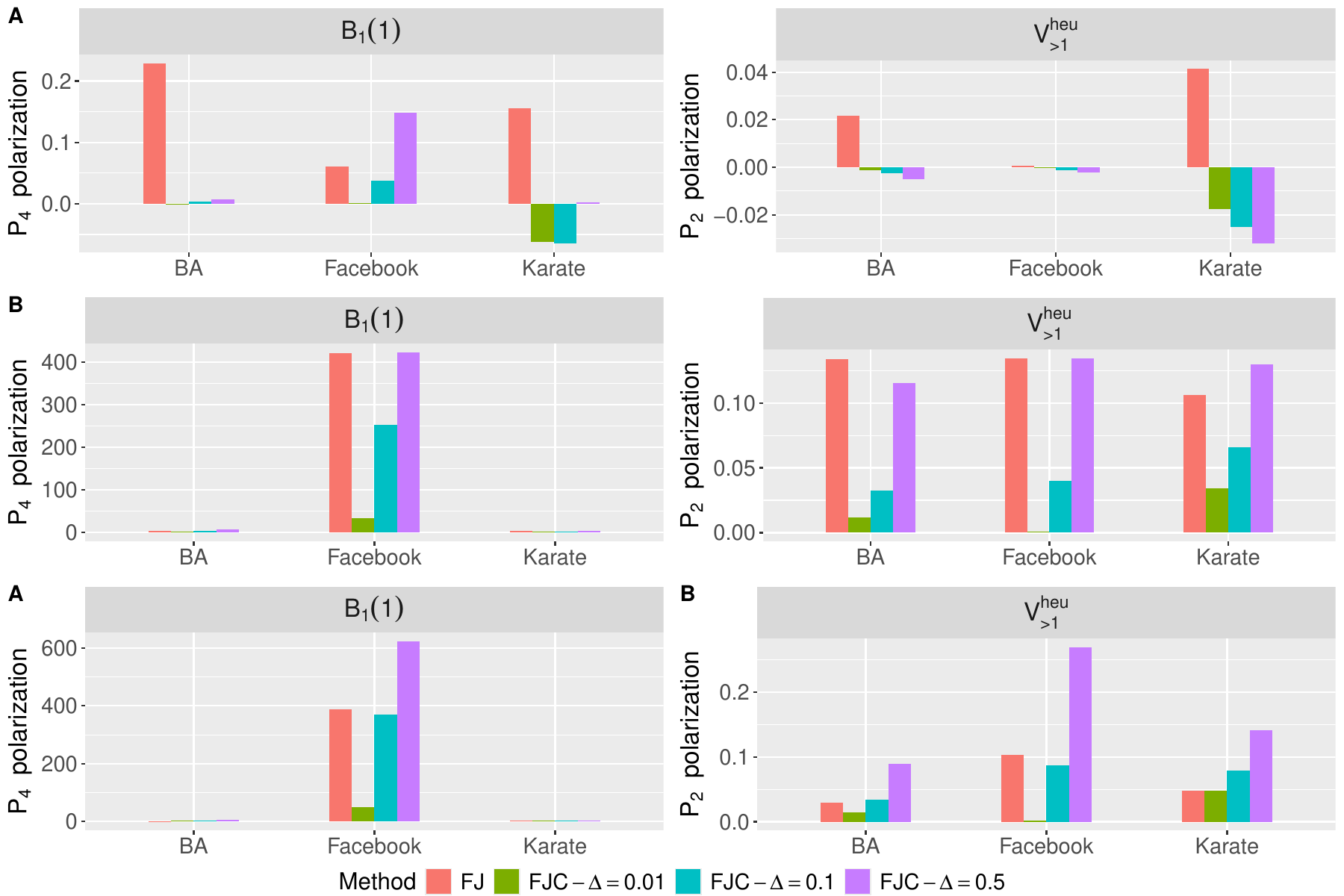}
 \caption{Polarization of FJ and FJC (with different $\theta$ values) for (A) $\lambda_i \sim \mathrm{P}_i$, (B) $\lambda_i \sim \mathrm{P}_i^{-1}$, (C) $\lambda_i = 0.6$. }
 \label{fig:polarization_synt}
 \vspace{-15pt}
\end{figure}

Figure~\ref{fig:polarization_synt} shows how opinion dynamics under FJC affect polarization compared to the FJ model, measured as the change in polarization (Equation~\eqref{eq:polarization}) between final and initial opinions. We report results for each susceptibility setting—proportional (A), inversely proportional (B), and uniform (C)—and for each $\theta$ value (differently colored bars for FJC). 
We show the polarization metric maximized by each polarizing vector in the FJ model: $P_4$ for $\bm{u}_{B_1(1)}$ and $P_2$ for the others. 
We first consider FJC with $\theta = 0.5$  (purple bars). When susceptibility is uniform (panel C), FJC consistently yields higher polarization than FJ, as central nodes remain influential and pull others toward their opinions. 
A similar, though less pronounced, pattern appears with inverse-centrality susceptibility (panel B), since less central nodes become more vulnerable under FJC. An exception arises for $\bm{u}_{{V_{>1}}^{heu}}$ in the BA network, where polarization decreases because the parameter $\theta$ limits opinion propagation, as discussed in the previous section.
Interestingly, when susceptibility is proportional to centrality (panel A), FJC often reduces polarization compared to FJ. Here, central nodes are nearly immovable, and peripheral nodes, unable to influence them, shift closer to reduce cognitive dissonance—flattening the opinion landscape. An exception occurs in the Facebook network with $\bm{u}_{B_1(1)}$, where the initial opinion aligns with a central node that remains firm, attracting others and increasing polarization.
For $\theta = 0.01$ and $\theta = 0.1$ (green and blue bars), the cascade effect is weak due to limited reposting, resulting in polarization trends similar to those at higher $\theta$ but with lower magnitude. In some cases, the effect even reverses, leading to depolarization—primarily driven by the few nodes directly connected to the sources, which are the only ones likely to update their opinions. For instance, in the Karate network with proportional susceptibility and the vector $\bm{u}_{B_1(1)}$, most neighbors of the sources have low susceptibility and thus remain unchanged. The only significant shift comes from the source node itself, which lowers its opinion from 1, reducing overall polarization as measured by $P_4$.

\section{Conclusions}

This study explores the dynamics of opinion formation and dissemination within online social networks, particularly under the influence of information cascades. Specifically, the paper proposes the Friedkin-Johnsen on Cascade model (FJC), which integrates opinion dynamics and information cascades. To the best of our knowledge, this is the first model that allows for the investigation of how content dissemination can affect opinion formation. 
We validated the FJC model, utilising a real cascade dataset. Our results demonstrate that the FJC model can reliably approximate final opinions accurately, even when many of the underlying assumptions of the model are relaxed or only partially met.
We evaluate the model using both synthetic and real social networks and highlight that the cascades reinforce the strength of central nodes in influencing opinions. These central nodes tend to be less vulnerable to opposing opinions, making it unlikely for them to change their views even when pressured by a large majority. This occurs when content is free to circulate within the network. Conversely, when participation is low, content cannot go viral, disrupting the opinion dynamics. 
The model builds upon certain assumptions regarding content dissemination and opinion evolution.

\vspace{-10pt}

\bibliography{references.bib}{}

@article{biondi2023dynamics,
  title={Dynamics of opinion polarization},
  author={Biondi, Elisabetta and Boldrini, Chiara and Passarella, Andrea and Conti, Marco},
  journal={IEEE TSMC: Systems},
  year={2023},
  publisher={IEEE}
}

@article{frasca2013gossips,
  title={Gossips and prejudices: Ergodic randomized dynamics in social networks},
  author={Frasca, Paolo and Ravazzi, Chiara and Tempo, Roberto and Ishii, Hideaki},
  journal={IFAC Proceedings Volumes},
  volume={46},
  number={27},
  pages={212--219},
  year={2013},
  publisher={Elsevier}
}

@inproceedings{kempe2003maximizing,
  title={Maximizing the spread of influence through a social network},
  author={Kempe, David and Kleinberg, Jon and Tardos, {\'E}va},
  booktitle={Proceedings of the ninth ACM SIGKDD}, 
  pages={137--146},
  year={2003}
}

@article{moller2020explaining,
  title={Explaining online news engagement based on browsing behavior: Creatures of habit?},
  author={M{\"o}ller, Judith and Van De Velde, Robbert Nicolai and Merten, Lisa and Puschmann, Cornelius},
  journal={SSCR},
  volume={38},
  number={5},
  pages={616--632},
  year={2020},
  publisher={SAGE Publications Sage CA: Los Angeles, CA}
}

@article{matsa2018news,
  title={News use across social media platforms 2018},
  author={Matsa, Katerina Eva and Shearer, Elisa},
  journal={Pew Research Center},
  volume={10},
  year={2018}
}

@article{friedkin1990social,
  title={Social influence and opinions},
  author={Friedkin, Noah E and Johnsen, Eugene C},
  journal={Journal of Mathematical Sociology},
  volume={15},
  number={3-4},
  pages={193--206},
  year={1990},
  publisher={Taylor \& Francis}
}

@inproceedings{Gionis2013,
	Author = {Gionis, Aristides and Terzi, Evimaria and Tsaparas, Panayiotis},
	Booktitle = {SIAM 2013},
	Doi = {10.1137/1.9781611972832.43},
	Isbn = {978-1-61197-262-7},
	Pages = {387--395},
	Publisher = {SIAM},
	Title = {{Opinion maximization in social networks}},
	Year = {2013},}

@article{Bindel2015,
	Author = {Bindel, David and Kleinberg, Jon and Oren, Sigal},
	Doi = {10.1016/J.GEB.2014.06.004},
	Issn = {0899-8256},
	Journal = {Games and Economic Behavior},
	Pages = {248--265},
	Publisher = {Academic Press},
	Title = {{How bad is forming your own opinion?}},
	Volume = {92},
	Year = {2015}}

@inproceedings{Musco2018,
	Author = {Musco, Cameron and Musco, Christopher and Tsourakakis, Charalampos E.},
	Booktitle = {Proceedings of the WWW '18},
	Doi = {10.1145/3178876.3186103},
	Isbn = {9781450356398},
	Pages = {369--378},
	Publisher = {ACM Press},
	Title = {{Minimizing Polarization and Disagreement in Social Networks}},
	Year = {2018}}

@inproceedings{Matakos2018,
	Author = {Matakos, Antonis and Gionis, Aristides},
	Booktitle = {ICDM 2018},
	Doi = {10.1109/ICDM.2018.00048},
	Isbn = {978-1-5386-9159-5},
	Pages = {327--336},
	Publisher = {IEEE},
	Title = {{Tell me Something My Friends do not Know: Diversity Maximization in Social Networks}},
	Year = {2018}}

@inproceedings{Chen2018,
	Author = {Chen, Xi and Lijffijt, Jefrey and {De Bie}, Tijl},
	Booktitle = {Proceedings of ACM KDD},
	Doi = {10.1145/3219819.3220074},
	Isbn = {9781450355520},
	Pages = {1197--1205},
	Publisher = {ACM Press},
	Title = {{Quantifying and Minimizing Risk of Conflict in Social Networks}},
	Year = {2018}}

@article{proskurnikov2017tutorial,
  title={{A tutorial on modeling and analysis of dynamic social networks. Part I}},
  author={Proskurnikov, Anton V and Tempo, Roberto},
  journal={Annual Reviews in Control},
  volume={43},
  pages={65--79},
  year={2017},
  publisher={Elsevier}
}

@inproceedings{albi2015optimal,
  title={On the optimal control of opinion dynamics on evolving networks},
  author={Albi, Giacomo and Pareschi, Lorenzo and Zanella, Mattia},
  booktitle={IFIP Conference on System Modeling and Optimization},
  pages={58--67},
  year={2015},
  organization={Springer}
}

@article{proskurnikov2018dynamics,
  title={{Dynamics and structure of social networks from a systems and control viewpoint: A survey of Roberto Tempo’s contributions}},
  author={Proskurnikov, Anton V and Ravazzi, Chiara and Dabbene, Fabrizio},
  journal={OSNEM},
  volume={7},
  pages={45--59},
  year={2018},
  publisher={Elsevier}
}

@book{Friedkin2011,
  title={Social influence network theory: A sociological examination of small group dynamics},
  author={Friedkin, Noah E and Johnsen, Eugene C},
  volume={33},
  year={2011},
  publisher={Cambridge University Press}
}

@article{Friedkin2017,
author = {Friedkin, Noah E and Bullo, Francesco},
doi = {10.1073/pnas.1710603114},
issn = {1091-6490},
journal = {PNAS},
month = {oct},
number = {43},
pages = {11380--11385},
pmid = {29073060},
publisher = {National Academy of Sciences},
title = {{How truth wins in opinion dynamics along issue sequences.}},
volume = {114},
year = {2017}
}

@article{Friedkin2015,
author = {Friedkin, Noah E.},
doi = {10.1109/MCS.2015.2406655},
issn = {1066-033X},
journal = {IEEE Control Systems},
month = {jun},
number = {3},
pages = {40--51},
title = {{The Problem of Social Control and Coordination of Complex Systems in Sociology: A Look at the Community Cleavage Problem}},
volume = {35},
year = {2015}
}

@inproceedings{Tu2023,
   author = {Sijing Tu and Stefan Neumann and Aristides Gionis},
   city = {New York, NY, USA},
   doi = {10.1145/3580305.3599255},
   isbn = {9798400701030},
   booktitle = {Proceedings of ACM SIGKDD},
   month = {8},
   pages = {2201-2210},
   publisher = {ACM},
   title = {Adversaries with Limited Information in the Friedkin-Johnsen Model},
   year = {2023},
}

@article{Sun2023,
   author = {Haoxin Sun and Zhongzhi Zhang},
   doi = {10.1609/aaai.v37i4.25585},
   issn = {2374-3468},
   issue = {4},
   journal = {Proceedings of the AAAI},
   month = {6},
   pages = {4623-4632},
   title = {Opinion Optimization in Directed Social Networks},
   volume = {37},
   year = {2023},
}

@article{abebe2021opinion,
  title={Opinion dynamics optimization by varying susceptibility to persuasion via non-convex local search},
  author={Abebe, Rediet and Chan, T-H HUBERT and Kleinberg, Jon and Liang, Zhibin and Parkes, David and Sozio, Mauro and Tsourakakis, Charalampos E},
  journal={ACM TKDD},
  volume={16},
  number={2},
  pages={1--34},
  year={2021},
  publisher={ACM New York, NY}
}

@article{chitra2019understanding,
  title={Understanding filter bubbles and polarization in social networks},
  author={Chitra, Uthsav and Musco, Christopher},
  journal={arXiv preprint arXiv:1906.08772},
  year={2019}
}

@inproceedings{Bhalla2023,
   author = {Nikita Bhalla and Adam Lechowicz and Cameron Musco},
   city = {New York, NY, USA},
   doi = {10.1145/3539597.3570442},
   isbn = {9781450394079},
   booktile = {Proceedings of WSDM'23},
   month = {2},
   pages = {6-14},
   publisher = {ACM},
   title = {Local Edge Dynamics and Opinion Polarization},
   year = {2023},
}

@inproceedings{bilo2022general,
   author = {Vittorio Bilò and Diodato Ferraioli and Cosimo Vinci},
   city = {California},
   doi = {10.24963/ijcai.2022/13},
   isbn = {978-1-956792-00-3},
   booktitle = {Proceedings of IJCAI},
   month = {7},
   pages = {88-94},
   title = {General Opinion Formation Games with Social Group Membership},
   year = {2022},
}

@inproceedings{Razaq2023,
   author = {Muhammad Ahsan Razaq and Claudio Altafini},
   doi = {10.1109/CDC49753.2023.10383479},
   isbn = {979-8-3503-0124-3},
   booktitle = {IEEE CDC'23},
   month = {12},
   pages = {491-496},
   title = {Propagation of Stubborn Opinions on Signed Graphs},
   year = {2023},
}

@inproceedings{disaro2023extension,
   author = {Giorgia Disarò and Maria Elena Valcher},
   doi = {10.1109/CDC49753.2023.10383235},
   isbn = {979-8-3503-0124-3},
   booktitle = {IEEE CDC'23},
   month = {12},
   pages = {3384-3389},
   title = {On an Extension of the Friedkin-Johnsen Model: The Effects of a Homophily-Based Influence Matrix},
   year = {2023},
}

@article{guille2013information,
  title={Information diffusion in online social networks: A survey},
  author={Guille, Adrien and Hacid, Hakim and Favre, Cecile and Zighed, Djamel A},
  journal={ACM Sigmod Record},
  volume={42},
  number={2},
  pages={17--28},
  year={2013},
  publisher={ACM New York, NY, USA}
}

@article{barbieri2013topic,
  title={Topic-aware social influence propagation models},
  author={Barbieri, Nicola and Bonchi, Francesco and Manco, Giuseppe},
  journal={KAIS},
  volume={37},
  pages={555--584},
  year={2013},
  publisher={Springer}
}

@article{aslay2016revenue,
  title={Revenue maximization in incentivized social advertising},
  author={Aslay, Cigdem and Bonchi, Francesco and Lakshmanan, Laks VS and Lu, Wei},
  journal={arXiv preprint arXiv:1612.00531},
  year={2016}
}

@inproceedings{chen2011influence,
  title={Influence maximization in social networks when negative opinions may emerge and propagate},
  author={Chen, Wei and Collins, Alex and Cummings, Rachel and Ke, Te and Liu, Zhenming and Rincon, David and Sun, Xiaorui and Wang, Yajun and Wei, Wei and Yuan, Yifei},
  booktitle={Proceedings of SDM'11},
  pages={379--390},
  year={2011},
  organization={SIAM}
}

@inproceedings{gayraud2015diffusion,
  title={Diffusion maximization in evolving social networks},
  author={Gayraud, Nathalie TH and Pitoura, Evaggelia and Tsaparas, Panayiotis},
  booktitle={ACM COSN},
  pages={125--135},
  year={2015}
}

@inproceedings{tu2022viral,
  title={A viral marketing-based model for opinion dynamics in online social networks},
  author={Tu, Sijing and Neumann, Stefan},
  booktitle={Proceedings of the ACM Web Conference},
  pages={1570--1578},
  year={2022}
}

@article{degroot1974reaching,
  title={Reaching a consensus},
  author={DeGroot, Morris H},
  journal={Journal of the American Statistical Association},
  volume={69},
  number={345},
  pages={118--121},
  year={1974},
  publisher={Taylor \& Francis Group}
}

@article{french1956formal,
  title={A formal theory of social power.},
  author={French Jr, John RP},
  journal={Psychological review},
  volume={63},
  number={3},
  pages={181},
  year={1956},
  publisher={American Psychological Association}
}

@inproceedings{Das2013,
  title={Debiasing social wisdom},
  author={Das, Abhimanyu and Gollapudi, Sreenivas and Panigrahy, Rina and Salek, Mahyar},
  booktitle={Proceedings of the 19th ACM KDD},
  pages={500--508},
  year={2013}
}

@book{easley2010networks,
  title={Networks, Crowds, and Markets: Reasoning about a Highly Connected World},
  author={Easley, David and Kleinberg, Jon},
  year={2010},
  publisher={Cambridge University Press}
}

@book{barabasi2016network,
  author = {Barabási, Albert-László},
  publisher = {Cambridge University Press},
  title = {Network science},
  year = 2016
}

@article{bernardo2021achieving,
  title={Achieving consensus in multilateral international negotiations: The case study of the 2015 Paris Agreement on climate change},
  author={Bernardo, Carmela and Wang, Lingfei and Vasca, Francesco and Hong, Yiguang and Shi, Guodong and Altafini, Claudio},
  journal={Science Advances},
  volume={7},
  number={51},
  pages={eabg8068},
  year={2021},
  publisher={American Association for the Advancement of Science}
}

@incollection{peralta2025opinion,
  title={Opinion dynamics in social networks: From models to data},
  author={Peralta, Antonio F and Kert{\'e}sz, J{\'a}nos and I{\~n}iguez, Gerardo},
  booktitle={Handbook of Computational Social Science},
  pages={384--406},
  year={2025},
  publisher={Edward Elgar Publishing Limited}
}

@article{zhou2021survey,
  title={A survey of information cascade analysis: Models, predictions, and recent advances},
  author={Zhou, Fan and Xu, Xovee and Trajcevski, Goce and Zhang, Kunpeng},
  journal={ACM Computing Surveys (CSUR)},
  volume={54},
  number={2},
  pages={1--36},
  year={2021},
  publisher={ACM New York, NY, USA}
}

@article{zachary1977information,
  title={An information flow model for conflict and fission in small groups},
  author={Zachary, Wayne W},
  journal={JAR},
  volume={33},
  number={4},
  pages={452--473},
  year={1977},
  publisher={University of New Mexico}
}

@inproceedings{leskovec2012learning,
  title={Learning to discover social circles in ego networks},
  author={Leskovec, Jure and Mcauley, Julian J},
  booktitle={NeurIPS'12},
  pages={539--547},
  year={2012}
}

@article{bhowmick2019temporal,
  title={Temporal sequence of retweets help to detect influential nodes in social networks},
  author={Bhowmick, Ayan Kumar and Gueuning, Martin and Delvenne, Jean-Charles and Lambiotte, Renaud and Mitra, Bivas},
  journal={IEEE TCSS},
  volume={6},
  number={3},
  pages={441--455},
  year={2019},
  publisher={IEEE}
}

@article{watel2016practical,
  title={A practical greedy approximation for the directed steiner tree problem},
  author={Watel, Dimitri and Weisser, Marc-Antoine},
  journal={JOCO},
  volume={32},
  number={4},
  pages={1327--1370},
  year={2016},
  publisher={Springer}
}

@inproceedings{tang2015influence,
  title={Influence maximization in near-linear time: A martingale approach},
  author={Tang, Youze and Shi, Yanchen and Xiao, Xiaokui},
  booktitle={ACM SIGMOD},
  pages={1539--1554},
  year={2015}
}

@article{brooks2025opinion,
  title={An “opinion reproduction number” for infodemics in a bounded-confidence content-spreading process on networks},
  author={Brooks, Heather Z and Porter, Mason A},
  journal={Chaos},
  volume={35},
  number={1},
  year={2025},
  publisher={AIP Publishing}
}

@inproceedings{Romero_Meeder_Kleinberg_2011,  
title={Differences in the mechanics of information diffusion across topics: idioms, political hashtags, and complex contagion on twitter},
DOI={10.1145/1963405.1963503}, 
booktitle={ACM WWW}, 
author={Romero, Daniel M. and Meeder, Brendan and Kleinberg, Jon},
year={2011}, pages={695–704} }

@inproceedings{Hodas_Lerman_2012,
  title={How visibility and divided attention constrain social contagion},
  author={Hodas, Nathan Oken and Lerman, Kristina},
  booktitle={Proceedings of PASSAT-SocialCom},
  pages={249--257},
  year={2012},
  organization={IEEE}
}

@article{pfeffer2023just, title={Just Another Day on Twitter: A Complete 24 Hours of Twitter Data}, volume={17}, rights={Copyright (c) 2023 Association for the Advancement of Artificial Intelligence}, ISSN={2334-0770}, DOI={10.1609/icwsm.v17i1.22215}, abstractNote={At the end of October 2022, Elon Musk concluded his acquisition of Twitter. In the weeks and months before that, several questions were publicly discussed that were not only of interest to the platform’s future buyers, but also of high relevance to the Computational Social Science research community. For example, how many active users does the platform have? What percentage of accounts on the site are bots? And, what are the dominating topics and sub-topical spheres on the platform? In a globally coordinated effort of 80 scholars to shed light on these questions, and to offer a dataset that will equip other researchers to do the same, we have collected all 375 million tweets published within a 24-hour time period starting on September 21, 2022. To the best of our knowledge, this is the first complete 24-hour Twitter dataset that is available for the research community. With it, the present work aims to accomplish two goals. First, we seek to answer the aforementioned questions and provide descriptive metrics about Twitter that can serve as references for other researchers. Second, we create a baseline dataset for future research that can be used to study the potential impact of the platform’s ownership change.}, journal={Proceedings of ICWSM}, author={Pfeffer, Jürgen and Matter, Daniel and Jaidka, Kokil and Varol, Onur and Mashhadi, Afra and Lasser, Jana and Assenmacher, Dennis and Wu, Siqi and Yang, Diyi and Brantner, Cornelia and Romero, Daniel M. and Otterbacher, Jahna and Schwemmer, Carsten and Joseph, Kenneth and Garcia, David and Morstatter, Fred}, year={2023}, month=jun, pages={1073–1081}, language={en} }

@inproceedings{garimella2018quantifying,
  title={Quantifying controversy on social media},
  author={Garimella, Kiran and Morales, Guillermo D and Gionis, Aristides and Mathioudakis, Michael},
  booktitle={ICWSM'18},
  pages={22--32},
  year={2018}
}

@inproceedings{suarez2025cross,
  title={Cross-Partisan Interactions on Twitter},
  author={Suarez-Tangil, Guillermo},
  booktitle={Proceedings of ICWSM},
  year={2025}
}

@article{vosoughi2018spread,
  title={The spread of true and false news online},
  author={Vosoughi, Soroush and Roy, Deb and Aral, Sinan},
  journal={Science},
  volume={359},
  number={6380},
  pages={1146--1151},
  year={2018},
  publisher={American Association for the Advancement of Science}
}

@article{delvicario2016misinformation,
  title={The spreading of misinformation online},
  author={Del Vicario, Michela and Bessi, Alessandro and Zollo, Fabiana and Petroni, Fabio and Scala, Antonio and Caldarelli, Guido and Stanley, H Eugene and Quattrociocchi, Walter},
  journal={PNAS},
  volume={113},
  number={3},
  pages={554--559},
  year={2016},
  publisher={National Acad Sciences}
}

@article{tsfati2003media,
  title={Do people watch what they do not trust? Exploring the association between news media skepticism and exposure},
  author={Tsfati, Yariv and Cappella, Joseph N.},
  journal={Communication Research},
  volume={30},
  number={5},
  pages={504--529},
  year={2003},
  publisher={SAGE Publications}
}

@article{festinger1957theory,
  title={A theory of cognitive dissonance},
  author={Festinger, Leon},
  journal={Stanford Univ. Press},
  year={1957}
}

@article{scarpa2025polarization,
  title={Polarization in Political Rallies: A Markovian Agent-Based Model for Opinion Dynamics},
  author={Scarpa, Marco and Garofalo, Marco and Longo, Francesco and Serrano, Salvatore},
  journal={Algorithms},
  volume={18},
  number={6},
  pages={308},
  year={2025},
  publisher={MDPI}
}

@article{liu2024agent,
  title={Agent-based modelling of polarized news and opinion dynamics in social networks: a guidance-oriented approach},
  author={Liu, Shan and Wen, Hao},
  journal={Journal of Complex Networks},
  volume={12},
  number={4},
  pages={cnae028},
  year={2024},
  publisher={Oxford University Press}
}

@article{zanella2023kinetic,
  title={Kinetic models for epidemic dynamics in the presence of opinion polarization},
  author={Zanella, Mattia},
  journal={Bull. Math. Biol.},
  volume={85},
  number={5},
  pages={36},
  year={2023},
  publisher={Springer}
}
\bibliographystyle{IEEEtran}

\vfill

\end{document}